\documentclass[aps,pre,showpacs,groupedaddress]{revtex4}%PRE
\usepackage{amsfonts}
\usepackage{mathrsfs}
\usepackage{dcolumn}
\usepackage{graphicx}
\usepackage{amsmath}
\usepackage{amssymb}
\usepackage{amsbsy}

\newcommand{\beq}{\begin{equation}}
\newcommand{\eeq}{\end{equation}}
\newcommand{\ba}{\begin{align}}
\newcommand{\ea}{\end{align}}

\def\bs{\boldsymbol \sigma}

\begin{document}

\title{Clustering of neural codewords revealed by a first-order phase transition}

\author{Haiping Huang}
%\email{physhuang@gmail.com}
\address{RIKEN Brain Science Institute, Wako-shi, Saitama
351-0198, Japan}
\author{Taro Toyoizumi}
%\email{taro.toyoizumi@brain.riken.jp}
\affiliation{RIKEN Brain Science Institute, Wako-shi, Saitama
351-0198, Japan}
\date{\today}

\begin{abstract}
A network of neurons in the central nervous system collectively
  represents information by its spiking activity states. Typically
  observed states, i.e., codewords, occupy only a limited portion of the
  state space due to constraints imposed by network interactions. Geometrical organization of codewords in
  the state space, critical for neural
  information processing, is poorly understood due to its high dimensionality.
 Here, we explore the organization of neural codewords using retinal data by computing the entropy of codewords
as a function of Hamming distance from a particular reference codeword. Specifically, we report that the
retinal codewords in the state space are divided into multiple distinct clusters separated by entropy-gaps, and that
this structure is shared with well-known associative memory networks in a recallable phase. Our
analysis also elucidates a special nature of the all-silent state. The all-silent state is surrounded by
the densest cluster of codewords and located within a reachable distance from most codewords. This
codeword-space structure quantitatively predicts typical deviation of a state-trajectory from its initial
state. Altogether, our findings reveal a non-trivial heterogeneous structure of the codeword-space that
shapes information representation in a biological network.

%\tableofcontents{}
\end{abstract}

 \maketitle

%%%%%%%%%%%%%%%%%%%%%%%%%%%%%%%%%%%%%%%%%%%%%%%%%%%%%%%%%%%%%%%%%
\section{Introduction}
Recent advances in multi-electrode recording techniques allow simultaneous measurements of neural activity from
a large population of interacting neurons~\cite{Stev-2011,Yuste-2015}. A population of neurons encodes various information by its collective
spiking activity patterns, namely, neural codewords~\cite{spike-1997}. These codewords are passed and interpreted by a downstream circuit for further information processing. 
Characterizing the organization of the codewords is therefore critical for our
understanding of neural coding.   

To characterize the distribution of codewords, a maximum entropy model~\cite{Jaynes-1957} with pairwise interaction terms
has been fitted to neuroscience data~\cite{Bialek-2006,Tang-2008}. This model that fits the first two moments of activity statistics
was reported to characterize real data well in small groups of neurons. Importantly, these studies also suggest that codewords
are restricted due to neural interactions within a small subset of the state space, namely, the space composed of all possible combinations of each neuron's binary activity.
However, the geometrical organization of codewords is not
well understood.

Interestingly, the codewords of the well-known Hopfield network~\cite{Hopfield-1982} are also restricted within a small subset of state
space due to strong constraints imposed by interactions between neurons. The state space of the Hopfield network is organized into
multiple basins of attraction~\cite{Amit-1985}, with which a simple Glauber dynamics~\cite{Glauber-1963} can recall one of memorized
patterns hinted by a distorted initial pattern. This is the so-called associative memory~\cite{Hopfield-1982,Amit-1985}. Although both the neuroscience model described above
and the Hopfield network belong to the pairwise maximum entropy model, it remains largely
unknown if their codeword-spaces, composed of all codewords, share common features. Recent investigation of retinal activity data revealed multiple
local energy minima (LEM) in a fitted maximum entropy model~\cite{Tkacik-2014}. However, it does not provide how neural codewords are geometrically
organized because demonstration of the codeword-space structure entails consideration of all possible states.

The high dimensionality of the state space prevents an exhaustive search except in small networks, and standard
dimensionality-reduction techniques can easily abolish underlying structure by neglecting many relevant dimensions.
Hence, an efficient new technique is in need to visualize the neural codeword-space. One insight is that distance
between codewords is an important factor that constrains neural dynamics---previous experiments have shown that
state transitions are mostly restricted to neighboring codewords and nearby
codewords are known to encode similar information~\cite{Stopfer-2003,Mark-2010}. Based on this observation, we propose the distance-constrained statistical mechanics
analysis~\cite{Zecchina-2008,Huang-JPA2013} to concisely characterize the codeword-space structure
based on the distance from a reference codeword. In particular, we present an advanced mean-field framework that computes the entropy of codewords as a
function of Hamming distance from any reference state. By applying this technique to both the Hopfield network and retinal
data, we explore their codeword-space structures, i.e., whether codewords are divided into multiple clusters.

%%%%%%%%%%%%%%%%%%%%%%%%%%%%%%%%%%%%%%%%%%%%%%%%%%%%%%%%%%%%%%%%%%%
\begin{figure}[h!]
(a)    \includegraphics[bb=0 0 351 265,scale=0.65]{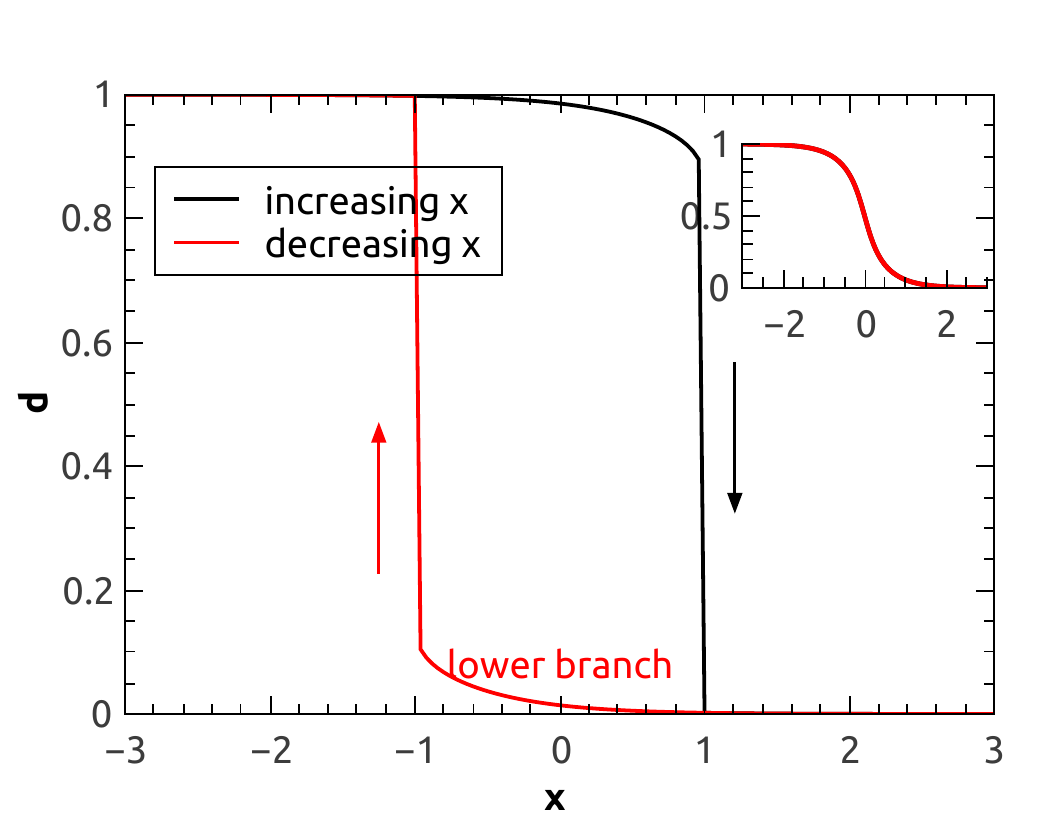}
     \hskip .1cm
 (b)    \includegraphics[bb=0 0 352 265,scale=0.65]{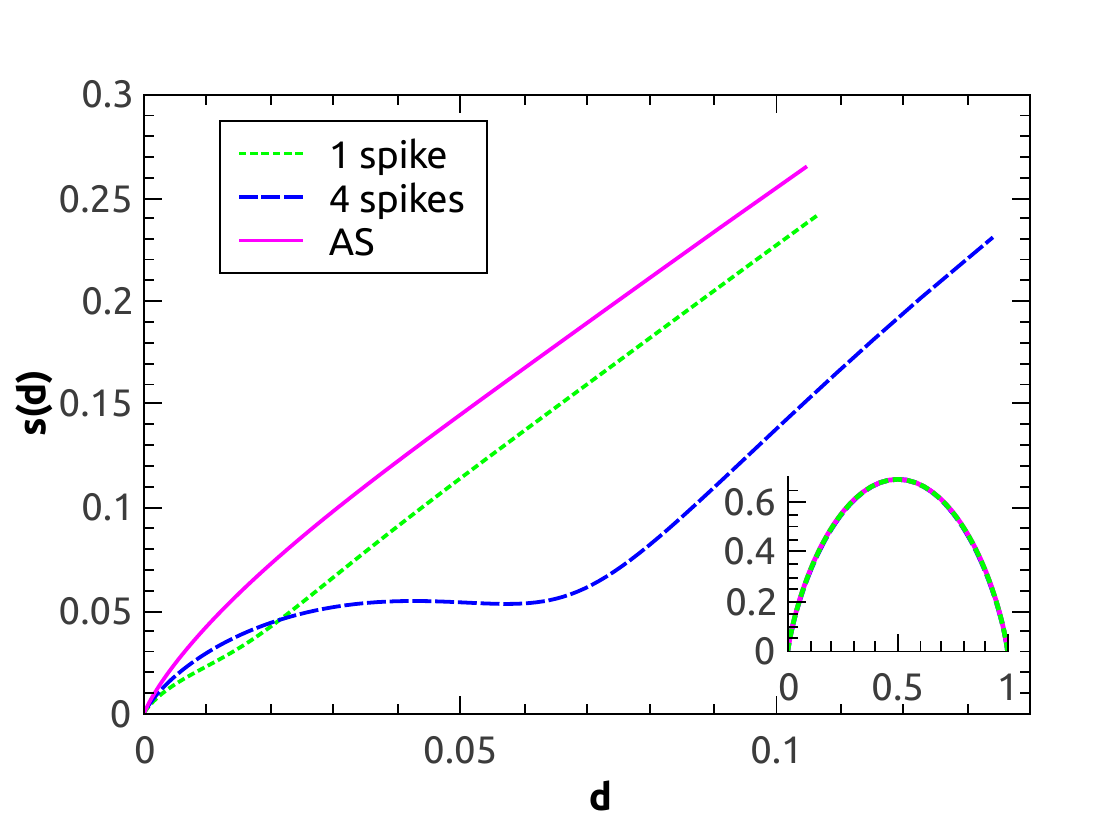}
     \vskip .1cm
  \caption{(Color online) Codeword organization of standard Hopfield model ($P=3$) with low spiking rate constraint of stored patterns.
  (a) a first order phase transition in Hamming distance $d$ when the coupling field $x$ is tuned. The reference pattern is the one with four spikes (see (b)). 
  The inset shows that the first order transition vanishes in the high temperature regime ($\beta=0.2$). (b) entropy per neuron as a function of Hamming distance from a 
  reference stored pattern with different spike-counts. The pattern with zero spike-count is named all-silent (AS) state. The curves correspond to the low-$d$ branch of the hysteresis loop (see (a)). The inset shows a trivial 
  entropy landscape identical for all references in high temperature regime.
     }\label{Hopf}
 \end{figure}

 %%%%%%%%%%%%%%%%%%%%%%%%%%%%%%%%%%%%%%%%%%%%%%%%%5
 \begin{figure}
\centering
     \includegraphics[bb=0 0 534 178,scale=0.85]{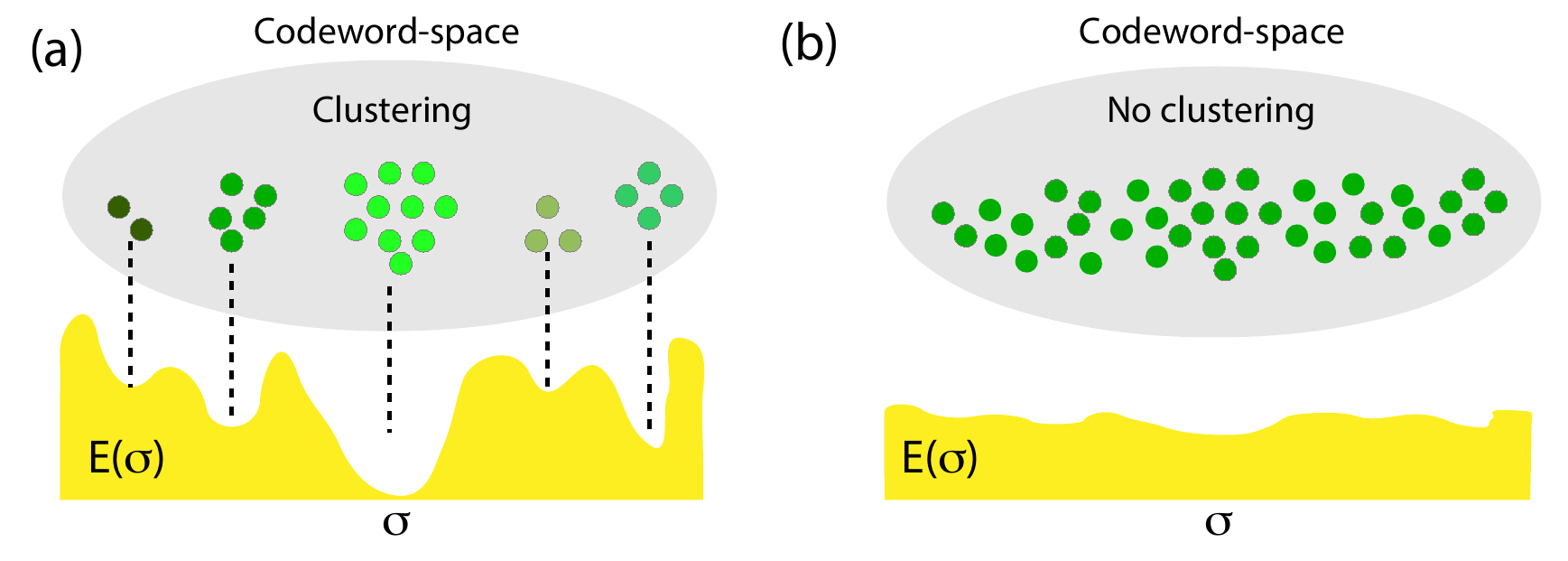}
  \caption{(Color online)
     Schematic illustration of two hypotheses on the organization of neural codewords in the state space. Each binary neural codeword ($\boldsymbol{\sigma}$) has an
     energy $E(\boldsymbol{\sigma})$. (a) When neural interactions are strong, neural codewords can be organized into multiple clusters in the state space.  (b) When the neural population is sufficiently noisy, a trivial structure (a single 
     cluster of neural codewords) is observed.}\label{lands}
 \end{figure}

\section{Results}
\label{Results}
\subsection{Distance-constrained statistical mechanics analysis}
We first introduce a statistical mechanics framework to characterize
codeword organization in the state space. Let $\sigma_i$ be binary activity of neuron
$i \,(i=1,\dots,N)$ and $\boldsymbol{\sigma}=(\sigma_1,\dots,\sigma_N)^T$ be a state
vector, representing population activity of $N$ neurons. Here
$\sigma_i=1$ indicates that neuron $i$ is active and $\sigma_i=-1$
indicates that neuron $i$ is silent. The symbol $^T$ represents
the transpose operation.

According to the maximum entropy principle~\cite{Jaynes-1957}, the activity state follows the Boltzmann distribution $P(\boldsymbol{\sigma})\propto\exp(-\beta E(\boldsymbol{\sigma}))$
where $\beta$ is the inverse temperature or neural reliability ($\beta=1$ unless otherwise indicated) and the energy $E(\boldsymbol{\sigma})=-\mathbf{h}^{T}\boldsymbol{\sigma}-
\frac{1}{2}\boldsymbol{\sigma}^{T}\mathbf{J}\boldsymbol{\sigma}$. $\mathbf{h}$ denotes a spiking bias vector and $\mathbf{J}$ a functional coupling matrix.
Geometrical organization of
codewords is studied by introducing a modified probability distribution
$P(\boldsymbol{\sigma})\propto\exp(-\beta E(\boldsymbol{\sigma})+x\boldsymbol{\sigma}^{T}\boldsymbol{\sigma}^{*})$, where  coupling field $x$ is introduced to control the overlap $\boldsymbol{\sigma}^{T}\boldsymbol{\sigma}^{*}$
between state $\boldsymbol{\sigma}$ and reference one $\boldsymbol{\sigma}^{*}$.
This perturbed probability measure gives the free energy per neuron defined by
\begin{eqnarray}
  \label{eq:f}
  f &\equiv& -\frac{1}{\beta N}\log\sum_{\boldsymbol{\sigma}}\exp(-\beta E(\boldsymbol{\sigma})+x\boldsymbol{\sigma}^{T}\boldsymbol{\sigma}^{*})\nonumber\\
  &=& -\frac{1}{\beta N}\iint d\epsilon dq  \exp(-N\beta f(\epsilon,q)),
\end{eqnarray}
where 
\begin{eqnarray}
  \label{eq:feq}
  \beta f(\epsilon,q)\equiv \beta \epsilon - x q - s(\epsilon,q)
\end{eqnarray}
is the energy- and overlap-dependent free energy that characterizes the
probability of states having energy $N\epsilon$ and overlap $Nq$, and $s(\epsilon,q)\equiv
(1/N)\log\sum_{\boldsymbol{\sigma}}\delta(\epsilon-E(\boldsymbol{\sigma})/N)\delta(q-\boldsymbol{\sigma}^{T}\boldsymbol{\sigma}^{*}/N))$
denotes entropy (log-number of states) per neuron with energy
$N\epsilon$ and overlap $Nq$. If the system-size $N$ is large,
the integral in Eq.~(\ref{eq:f}) is typically dominated by a
combination $(\epsilon,q)$ that minimizes $f(\epsilon,q)$, i.e.,
$f\approx \min_{\epsilon,q}f(\epsilon,q)$.

We compute $\epsilon$ and $q$ that minimize $f(\epsilon,q)$ by
applying the Bethe approximation~\cite{MM-2009} (see Methods). By recursively solving the mean field
equation, we estimate a local (or global) minimum of the free
energy and $(\epsilon,q)$ corresponding to this
minimum. Notably, these values of $\epsilon$ and $q$ characterize the
energy and overlap of typically observed states (namely codewords), respectively. Meanwhile, the entropy of codewords
$s(\epsilon,q)$ can be also computed according to
Eq.~(\ref{eq:feq}). We define Hamming distance $(N-\boldsymbol{\sigma}^T\boldsymbol{\sigma}^*)/2$ that counts how
many neurons have distinct activity in state $\boldsymbol{\sigma}$
and reference state $\boldsymbol{\sigma}^*$. The typical value of the overlap $q$ can be transformed to the typical value of Hamming distance per
neuron $d=(1-q)/2$. In the following sections, we omit
the $\epsilon$ dependency of the entropy and report it as a
function of $d$, i.e., $s(d)$.

%\subsection{The neural data shows a similar first-order transition to that in the standard Hopfield model}

\subsection{Clustering of codewords in the Hopfield model}
Using the mean field method, we first investigate the structure of
codeword-space in the Hopfield network~\cite{Amit-1985,Amit-1987}. In
this model, the coupling between neuron $i$ and $j$ is constructed as
$J_{ij}=1/N\sum_{\mu=1}^{P}\xi_i^{\mu}\xi_{j}^{\mu}$ for a network of
$N=60$ neurons, where $P=3$ random binary patterns (indexed by $\mu=1,\dots,P$) are stored.
In each pattern, stored activity $\xi_i^\mu$ of neuron $i$ takes $+1$ with probability
$r=0.0338$ and $-1$ with probability $1-r$. $r$ is chosen to fit the
activity level of retinal neurons we study in the next section. Note
that, in the Hopfield model, the neurons have zero spiking bias parameters
($\mathbf{h}=\mathbf{0}$). According to the previous section, we
compute the typical distance $d$ as we increase $x$ from $-3$ to $3$, and then decrease it from $3$ to $-3$ (Fig.~\ref{Hopf} (a)). More precisely,
after the convergence of the mean field equations at some $x$, we change $x$ by a
small amount and restart iteration from the previous fixed point (see
Methods). The reference state $\boldsymbol{\sigma}^*$ is set to one of the
stored patterns. Remarkably, we find a first-order phase transition
of $d$, characterized by the hysteresis loop (Fig.~\ref{Hopf} (a)). As we decrease $x$ from
high to low values, the typical distance suddenly jumps at around $x=-1.08$ from $d\approx0.16$ to
$d\approx0.97$, implying a non-trivial structure of the codeword-space.

In order to more directly visualize the non-trivial structure of the
codeword-space, we plot the entropy of codewords computed at
various distance $d$ away from each stored pattern. Only the entropy values corresponding to
 the low-$d$ branch of the hysteresis loop are shown in the figure. As shown in
 Fig.~\ref{Hopf} (b), each stored pattern has a dense core of codewords
around itself, which discontinuously falls off at some distance. This
indicates that codewords are organized into multiple clusters,
separated by non-codeword states (i.e., gaps). 
Among three stored patterns, the all-silent (AS) state has the largest core due to the low spiking rate constraint of stored
patterns (small $r$).

This clustering results from the attractor structure~\cite{Amit-1987} in the retrieval
phase of the model.
Within the hysteresis loop, there are two local minima of the free energy (Eq.~(\ref{eq:feq})) competing with each other. Low-$d$ minimum corresponds
to the nearby codewords of stored patterns ($\boldsymbol{\xi}$), while high-$d$ minimum corresponds to nearby codewords of corresponding reversed patterns ($-\boldsymbol{\xi}$). Thus each stored pattern has
distinct entropy landscape surrounding it.
The codeword-space clustering is necessary
for successful memory retrieval in the Hopfield network. In fact, in a high
temperature regime (non-recallable phase), the first order transition and the non-trivial entropy landscape are absent, as observed in the insets
of Fig.~\ref{Hopf}. All reference patterns display the same entropy landscape without entropy gaps, and thus the patterns can not be distinguished from each other.
In this non-recallable phase of the Hopfield network, there do exist multiple LEM (see Fig.~\ref{lands} (b))
under greedy descent dynamics (GDD, see Methods) on the energy surface, while
the codeword-space structure is trivial without entropy gaps.

\begin{figure}[h!]
(a)   \includegraphics[bb=0 0 351 265,scale=0.65]{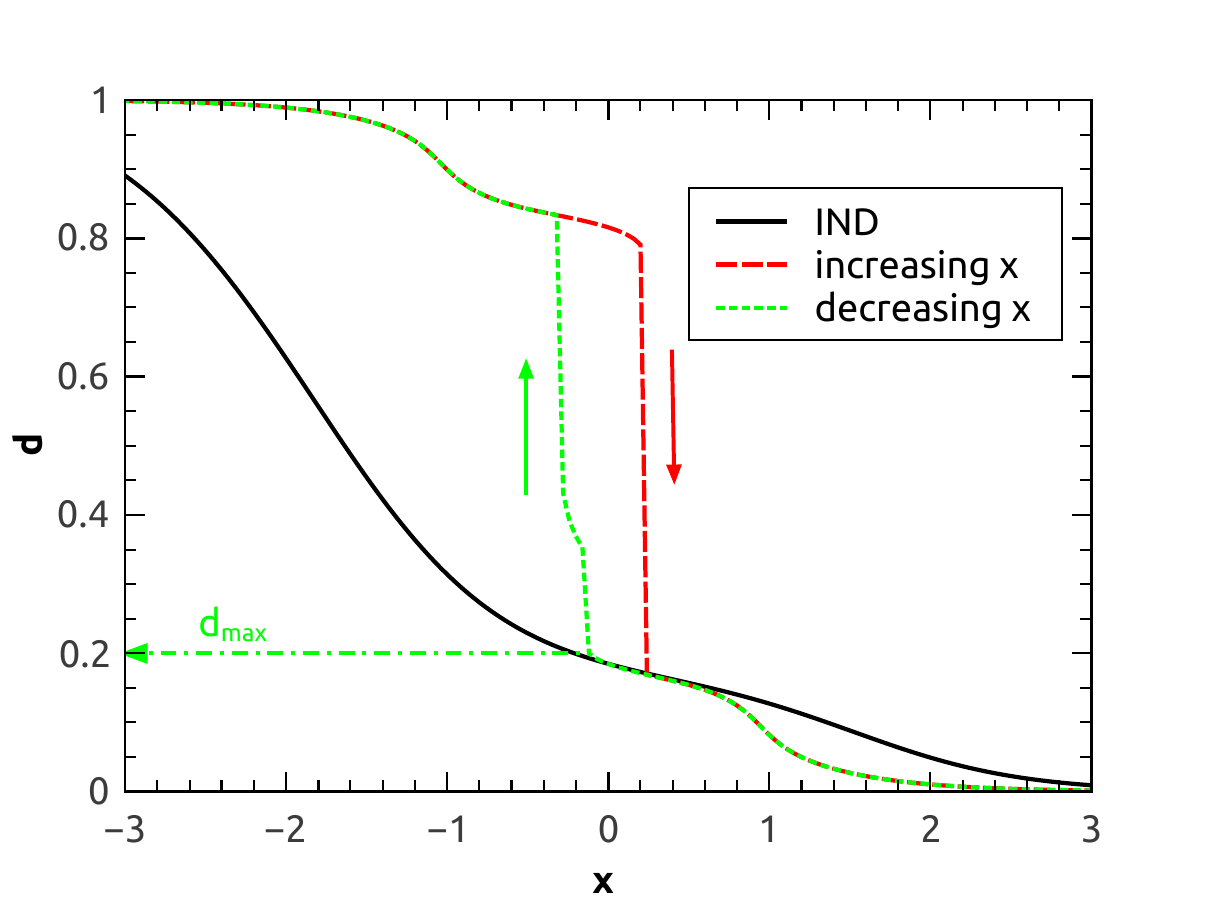}
     \hskip .1cm
 (b)     \includegraphics[bb=0 0 351 265,scale=0.65]{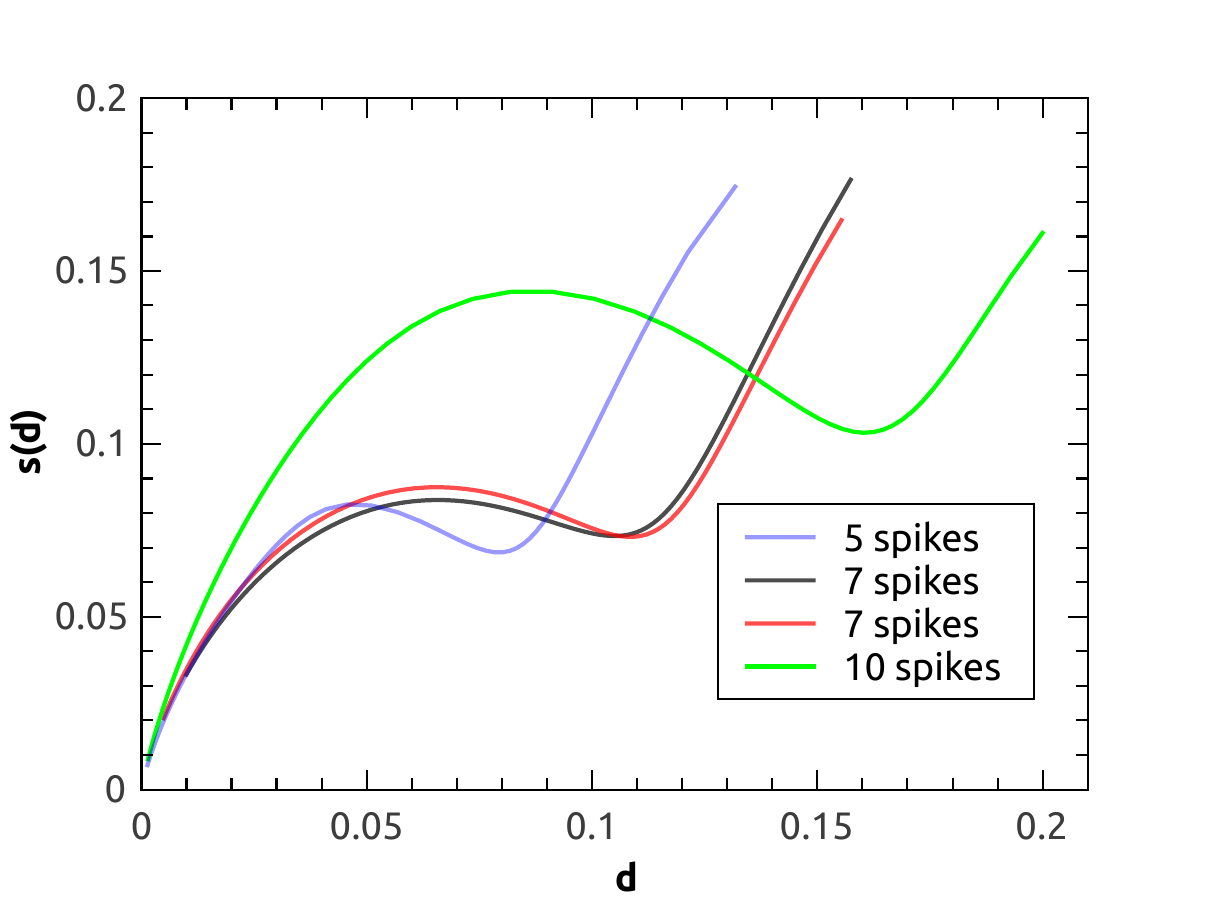}
     \vskip .1cm
     (c)    \includegraphics[bb=0 0 353 265,scale=0.65]{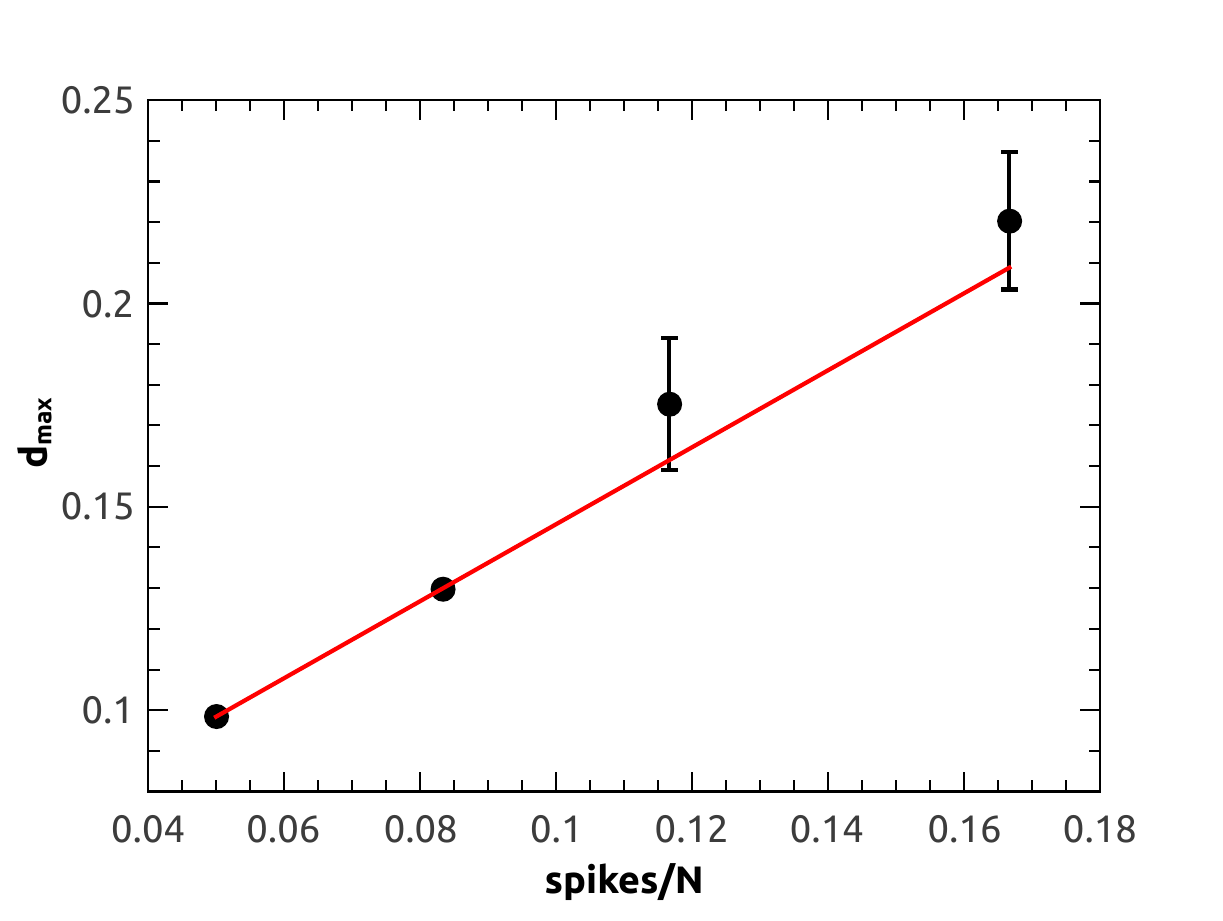}
     \hskip .1cm
 (d)    \includegraphics[bb=0 0 461 365,scale=0.5]{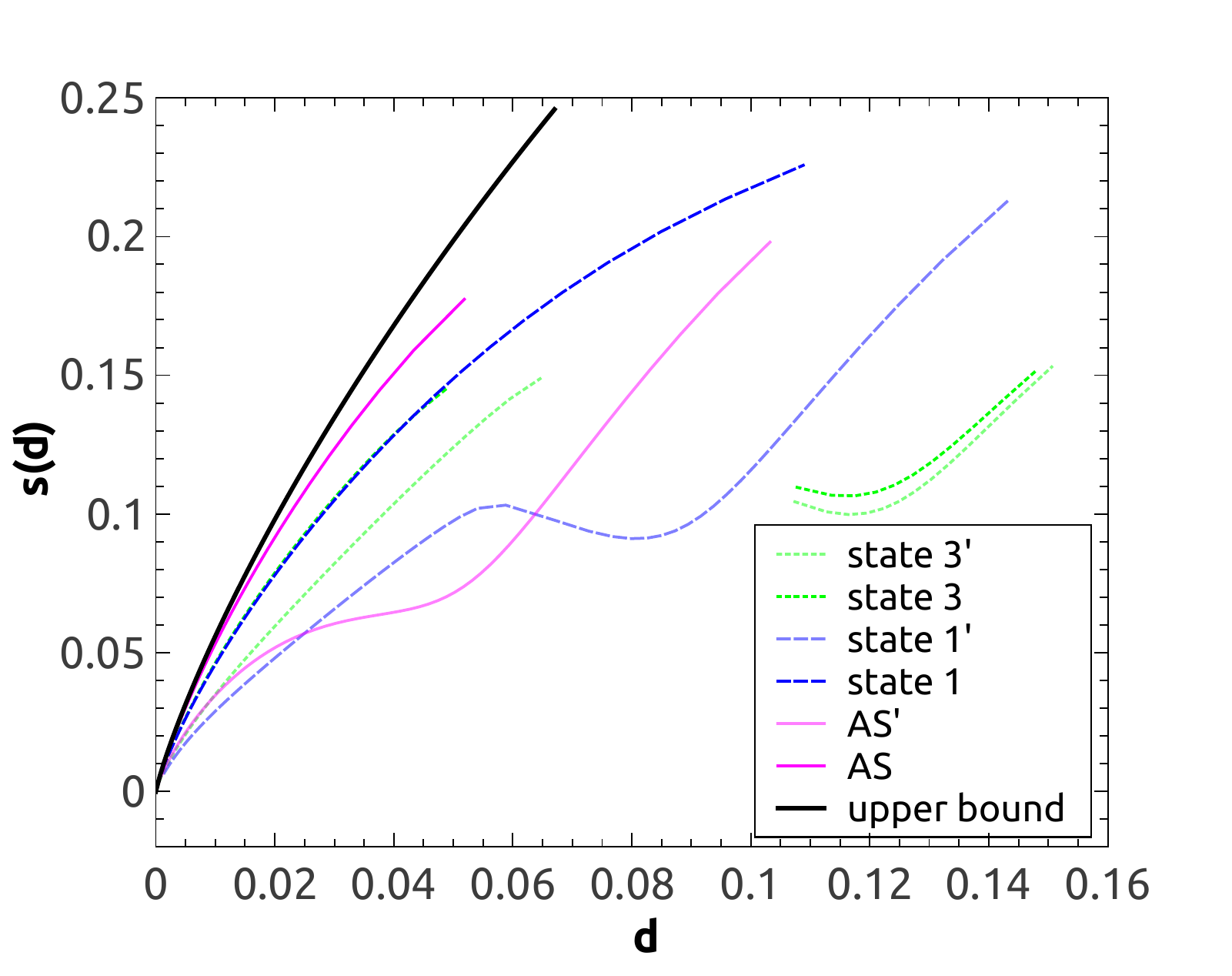}
     \vskip .1cm
  \caption{(Color online) Codeword organization of the neural data ($N=60$).
  (a) a first order phase transition in Hamming distance when the coupling field is tuned. The reference is a codeword of ten spike-counts. The neural codeword-space structure is shaped by the correlations in the neural spiking activity. The first-order
  transition disappears for independent model (IND). $d_{{\rm max}}$ defines the distance at which the low-$d$ branch in the hysteresis loop terminates.
  (b) distance-dependent
  entropy landscape from reference neural codewords of different spike-counts. (c) maximum distance $d_{{\rm max}}$ 
  versus spike-counts of the reference states (distance from AS state). Five references for each spike-count are randomly chosen. The line is a linear fit. (d) distance entropy from neural codewords (state $a'$) and their corresponding 
  LEM (state $a$). The corresponding LEM are identified by GDD. The random codeword limit is the upper bound. 
     }\label{retina}
 \end{figure}
 
 \subsection{Clustering of codewords in the retinal network}
The next important question is
how codewords of a real neural population are organized. To elucidate
this question, we analyze spiking activity data of populations of retinal ganglion cells under a repeated naturalistic movie stimulus~\cite{Marre-2012,Tkacik-2014}.
Although multiple LEM were previously found using
this data set~\cite{Tkacik-2014}, it is still unknown if the observed
network has clustering of codewords or not (Fig.~\ref{lands} (a)). We therefore characterize the geometrical organization of retinal codewords by applying the same
method as used in the Hopfield network. 
 
The neural spike trains in a population of $N$ neurons are binned with a $20$ ms temporal resolution to have $N$-dimensional spiking states $\boldsymbol{\sigma}$. Spiking bias $\mathbf{h}$ and functional coupling $\mathbf{J}$ are fitted to the spike train data to reproduce 
the mean activity and pairwise correlation of the data (see Methods). We choose randomly a network sample of the size $N=60$ from the neural data (the behavior reported below does not change qualitatively when
another sample is chosen, see supplementary Fig.~\ref{retina02}). Despite no clear similarity in the
connectivity structure to the Hopfiled network, the retinal network displays the first-order phase transition with a
hysteresis loop, qualitatively resembling the Hopfield model (Fig.~\ref{retina} (a)). This establishes that codewords of the retinal network are also
clustered. Furthermore, by constructing an independent maximum entropy model,
where only the mean activity is
fitted to the data with $\mathbf{J}=0$,
we show that the first-order phase transition disappears, indicating
that it is the non-trivial neural correlations that shape the clustering of codewords.

 %\subsection{The retinal network organizes its codeword-space into multiple distinct domains}
 
 Fig.~\ref{retina} (b) shows the entropy as a function of $d$ when neural codewords of different spike-counts are selected as references. 
 Again, only the entropy values corresponding to
 the low-$d$ branch of the hysteresis loop are shown in the figure. The high-$d$ branch is not biologically plausible, since
 the neural code is sparse. The entropy landscape is strongly dependent of the reference. In general, the higher spike-counts a neural codeword has, the larger distance its entropy curve extends
 over, enhancing the ability of the high spike-count codeword to come back to the sparse coding regime around the AS state. To quantify this property, Fig.~\ref{retina}
(c) plots the maximum distance $d_{{\rm max}}$ at which the low-$d$
branch in the hysteresis loop terminates as a function of spike-counts
of the reference codeword. We
 find that $d_{{\rm max}}$ increases linearly with the spike-counts (distance from the AS state) and the
estimated slope is $0.9462\pm0.0372$. The slope close to one is also
observed in another typical example (see supplementary Fig.~\ref{retina02}). Note that the distance to the AS state is typically smaller than $d_{{\rm max}}$. This implies that, 
 even if the neural codeword is far away from the AS state, it still has easy access to the sparse coding regime around the AS state within reasonable time, 
 which highlights the potential role of the AS state~\cite{Shimazaki-2015}.
 
 The AS state plays a special role here because the entropy curve from the AS reference state grows much more rapidly as a function of distance
than from the other codewords (Fig.~\ref{retina} (d)). Indeed, its growth is close to the upper bound given by the random codeword
limit ($s_{{\rm ub}}(d)=\frac{1}{N}\ln\binom{N}{Nd}$), in which every state is equally likely. This indicates that the AS state has the densest core of
codewords around it, which would facilitate frequent visits from other neural codewords (see supplementary Fig.~\ref{EvoD}).
 As previously observed~\cite{Tkacik-2014}, a large portion of neural patterns (about $94.25\%$ of $2000$ patterns) are observed to
 evolve to the AS state by following GDD (see Methods). 
 
 Fig.~\ref{retina} (d) reports the distance-dependent entropy landscape for some reference LEM codewords (e.g., state $a$) obtained by running the GDD method
 starting from corresponding reference non-LEM codewords (resp. state $a'$) (see the corresponding multidimensional scaling (MDS) map of LEM in supplementary Fig.~\ref{mdsLN} (a)). The result shows that each reference has a different
 landscape, and at small $d$, the entropy around a non-LEM codeword is typically smaller than that for
the corresponding LEM codeword. Moreover, for some states (e.g., $3$ and $3'$), there exist two continuous parts separated by a gap in the distance 
 entropy curve. We shall elaborate this phenomenon in the following section by studying a larger
population, where the effect becomes much more evident.  This shows another clear evidence for the clustering of neural
codewords.
 
% The non-trivial structure of the neural codeword-space stems from the (weakly) correlated neural responses.  This is confirmed by constructing an independent maximum
%entropy model that captures only mean spiking activity level in individual
%neurons. As shown in Fig.~\ref{retina} (a), the first order phase transition disappears, 
% resulting in a smooth entropy landscape. Thus, we conclude that it is the neural correlation that shapes the non-smooth energy
% (entropy) landscape, which might be helpful for efficient neural computation and sensory processing. 
 \begin{figure}
  (a) \includegraphics[bb=0 0 469 383,scale=0.45]{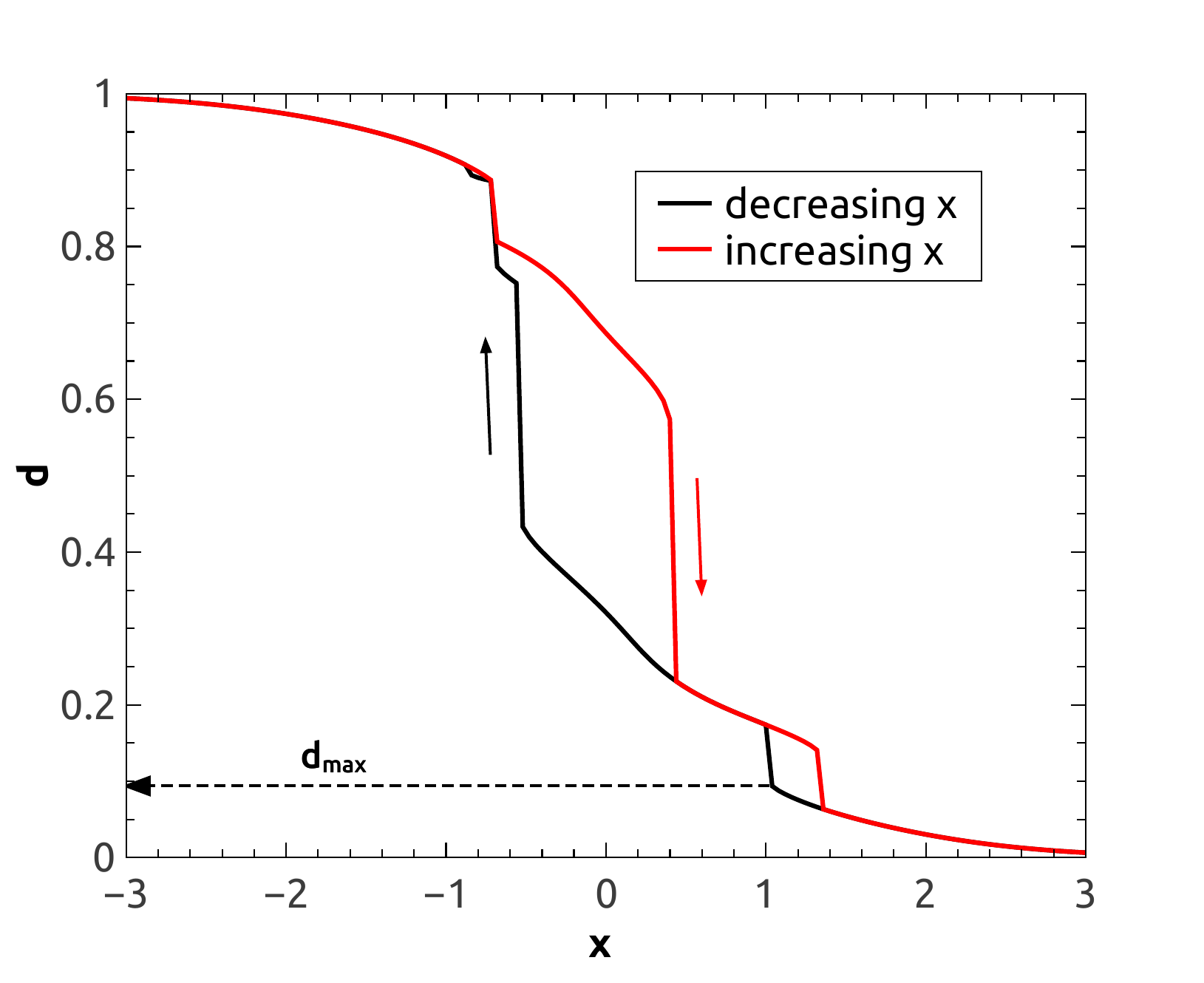}
   \hskip .1cm
    (b) \includegraphics[bb=0 0 351 265,scale=0.65]{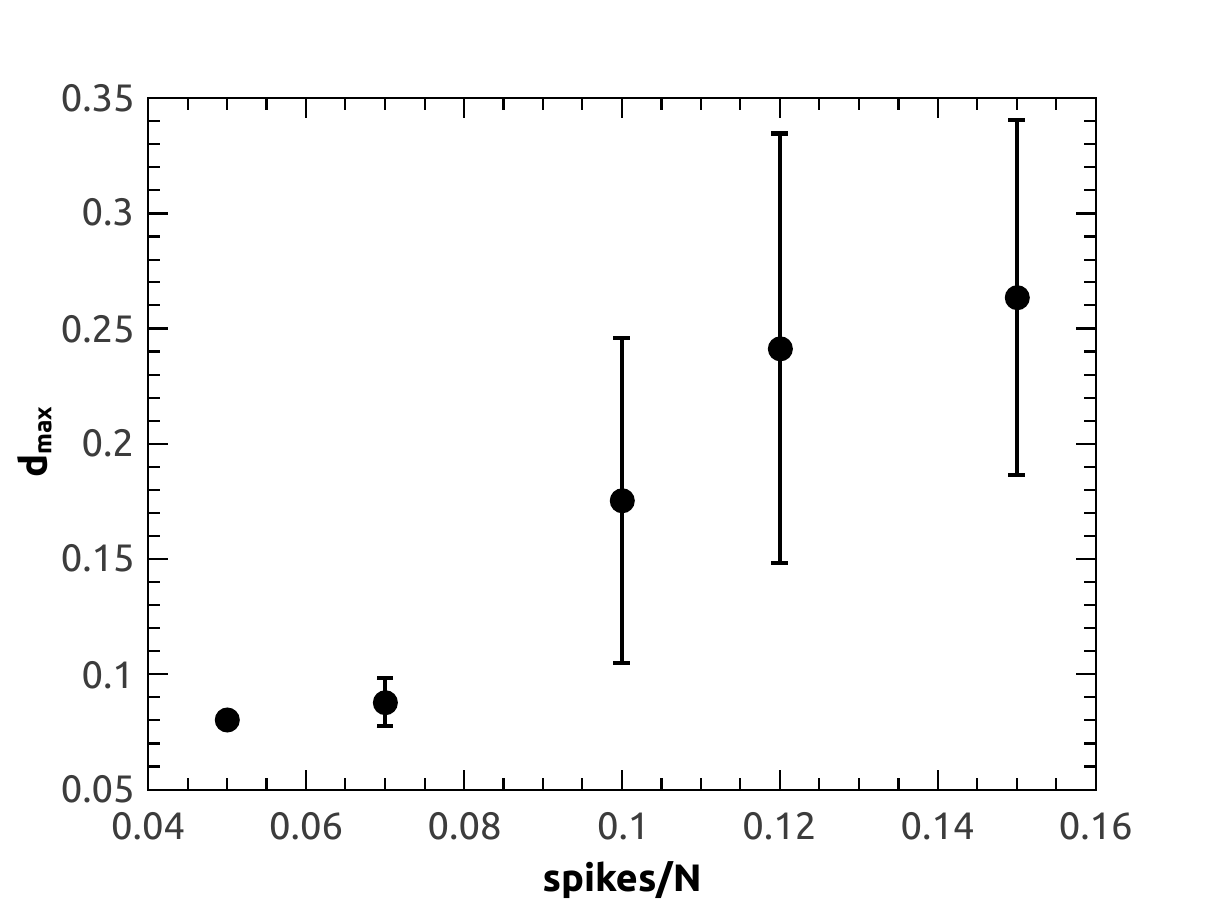}
 \vskip .1cm
    (c) \includegraphics[bb=0 0 504 326,scale=0.55]{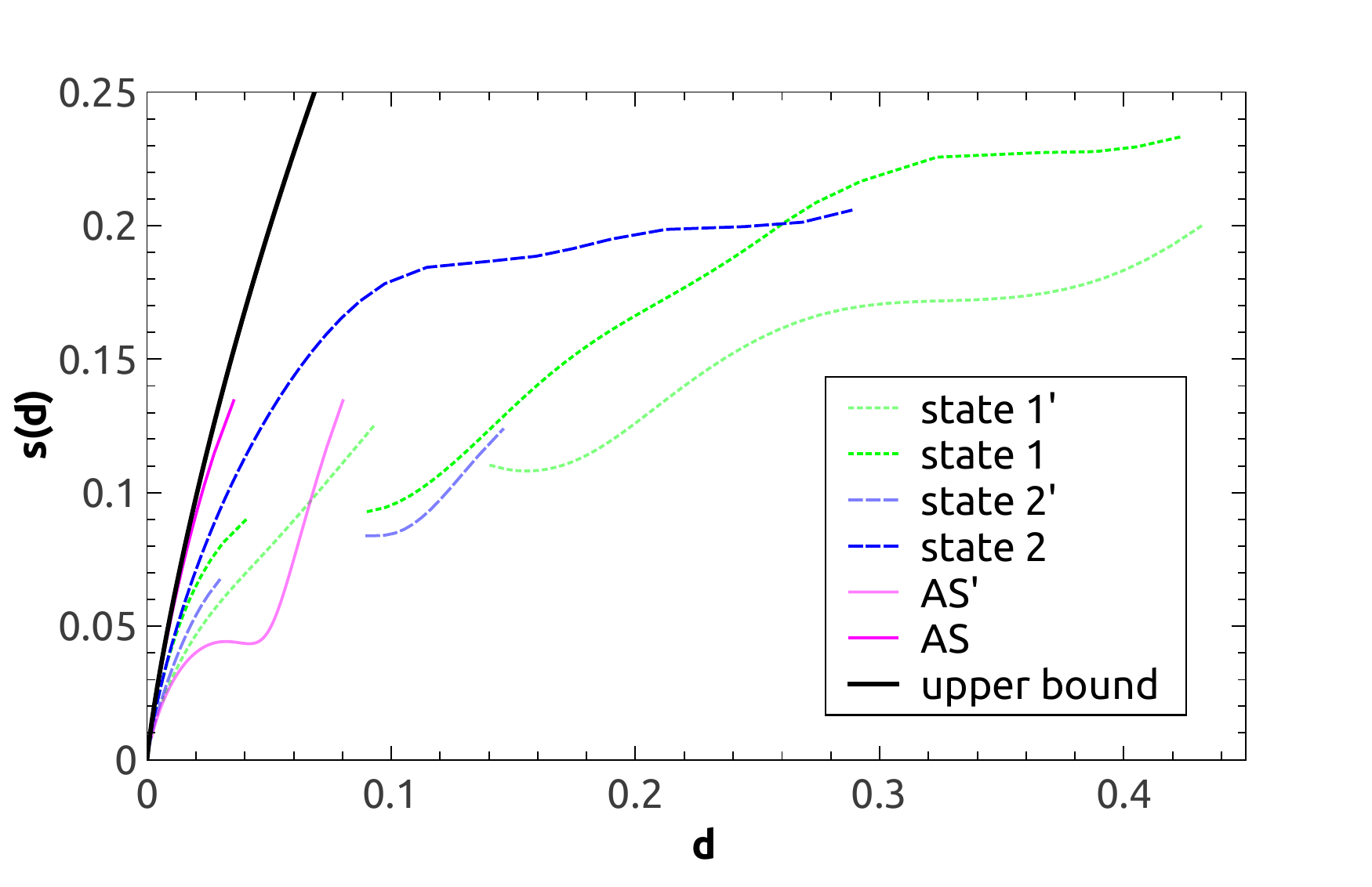}
     \vskip .1cm
  \caption{(Color online)
 Codeword organization of the neural data ($N=100$). (a) two hysteresis loops are observed. The reference is state $1'$ (see (c)). (b) maximum distance $d_{{\rm max}}$ at which the low-$d$ branch in the hysteresis loop terminates
  versus spike-counts of the reference (distance from AS state). Five references for each spike-count are randomly chosen. (c) entropy curve for different codewords and their assigned LEM. 
  Complex structure is observed for state $1$, $1'$ and $2'$.
     }\label{retinaLN}
 \end{figure}
 
 \subsection{More complicated structure observed for large neural populations}
 
 The property of the neural codeword shown above is still preserved when large populations of neurons are considered. In Fig.~\ref{retinaLN},
 we show the theoretical result computed on a network sample of $N=100$. As the network size grows, the number of LEM
 detected by GDD method also increases. Accordingly, the internal structure of the codewords becomes
 more complicated (a rough visualization is given by the MDS map, see supplementary Fig.~\ref{mdsLN} (b)). As shown in Fig.~\ref{retinaLN} (a), there exist two hysteresis loops separated 
 by another monostable branch (two curves for increasing and
decreasing $x$ coincide with each other). These two successive hysteresis loops naturally arise if there exist three deep
minima in the energy landscape, where sweeping $x$ shifts a dominant contribution from one to another. Fig.~\ref{retinaLN} (b) shows that
 $d_{{\rm max}}$ grows with spike-counts (distance from the AS state). The growth is likely nonlinear in this case,
perhaps induced by the complexity of the state space. The fraction of neural codewords that can reach the AS state
without in-between gaps reduces from the result of the previous section to about $76.05\%$.  Note that this number is
 still dominant compared to the reachability of other detected LEM. Again, the AS state has the densest surrounding core, characterized
by the rapid growth of the entropy with distance (Fig.~\ref{retinaLN} (c)). The entropy landscape surrounding the AS state does not have a second monostable branch
beyond the first entropy gap, except at a biologically implausible distance close to $1$. This might be because there
is no deep enough minima around the AS state. In contrast, the entropy landscape surrounding some other reference
codewords, e.g., state $1'$, exhibits a second monostable branch beyond the first entropy gap (see Fig.~\ref{retinaLN} (a)), likely indicating that
there is another deep minimum around them.

\begin{figure}
  (a) \includegraphics[bb=0 0 443 352,scale=0.5]{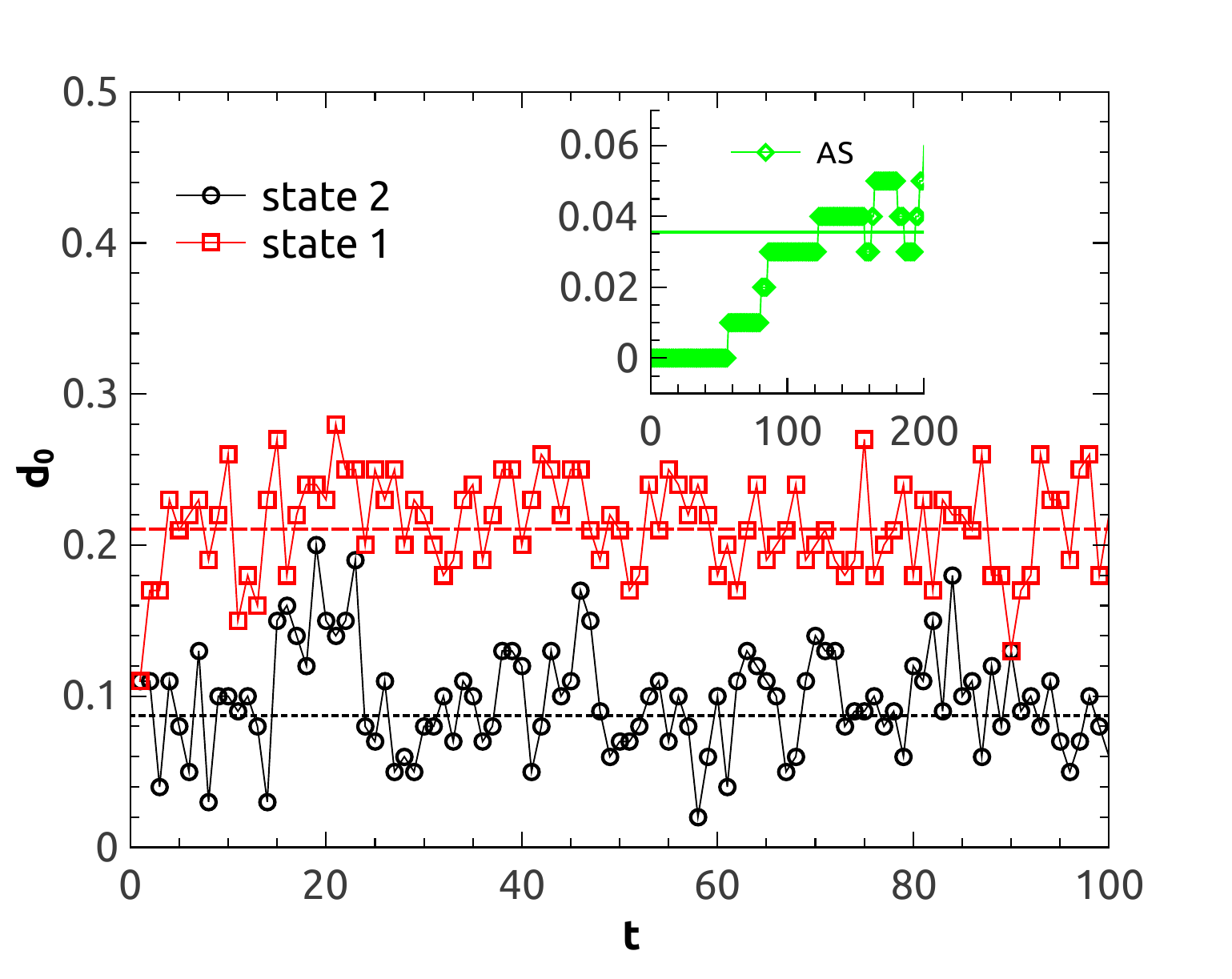}
   \hskip .1cm
    (b) \includegraphics[bb=0 0 340 265,scale=0.65]{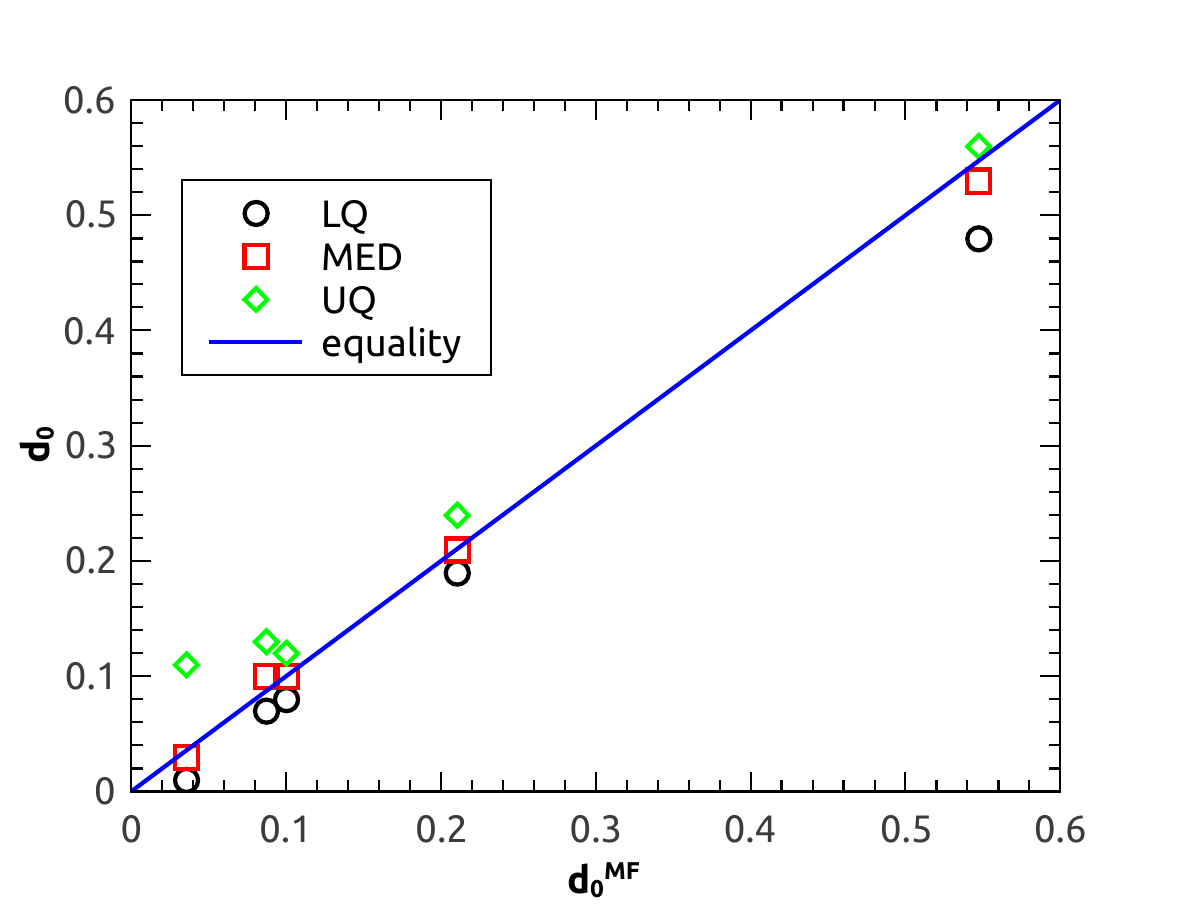}
 \vskip .1cm
  \caption{(Color online)
  Mean field theory predicts the plateau of the distance dynamics starting from LEM ($d_{0}(t)$). (a) typical trajectory observed for reference AS (inset), state 1 and 2 in simulations. The (solid, dashed, dotted) line is
  the theoretical prediction computed at $x=0$ for each reference. (b) the fluctuation plateau of $d_{0}(t)$ is predicted by the mean field theory ($d_{0}^{{\rm MF}}$). 
  Five trials from the 
  same reference are considered for each data point. Each trial lasts for $100$ steps. Each step corresponds to $N$ proposed flips. Note that in the inset of (a), one step corresponds
  to one possible flip. LQ: lower quartile; MED: median; UQ: upper
  quartile. 
     }\label{dynamics}
 \end{figure}

 To demonstrate the implication of the entropy landscape, we study how distance from a local energy minimum changes with time when the neural system explores the state
 space. We use the local dynamics rule characterized by
 the transition probability $w(\sigma_i\rightarrow-\sigma_i|H_i)=e^{-2\sigma_iH_i}$ where
 $H_i=h_i+\sum_{j}J_{ij}\sigma_j$ denotes the effective spiking bias of neuron $i$. Under this dynamics, states are sampled from the
 original distribution $P(\bs)\propto \exp(-\beta E(\bs))$. Note that the GDD rule to obtain LEM allows only monotonically decreasing energy on the energy surface.
 In contrast, the current dynamics rule allows the energy to increase occasionally. Sampled distance from a reference local energy minimum
 $\bs^*$ is denoted by
 $d_0(t)=(1-\boldsymbol{\sigma}^{T}(t)\boldsymbol{\sigma}^{*}/N)/2$
 where $t$ denotes the time step. The mean field prediction $d_0^{{\rm MF}}$ of a typical codeword-distance is given by setting $x=0$ and initializing the iteration
 equation (see Methods) at $\boldsymbol{\sigma}^{*}$. Note that $x=0$ corresponds to the case without distance-constraint, and thus takes into account all codewords in the cluster that $\boldsymbol{\sigma}^{*}$
 belongs to. As
 expected, this calculation predicts the fluctuation plateau of $d_0(t)$ close to the reference, as shown in Fig.~\ref{dynamics} (a) and (b). Note that 
 the local dynamics escapes fast from the AS state (see the inset of Fig.~\ref{dynamics} (a)), which may be related to its very small core (Fig.~\ref{retinaLN} (c)). The same qualitative behavior holds for the 
 smaller network ($N=60$, see supplementary Fig.~\ref{dynn60}) and when the neural dynamics is simulated starting from a non-LEM codeword. 
 
%%%%%%%%%%%%%%%%%%%%%%%%%%%%%%%%%%%%%%%%%%%%%%%%%%%%%%%%%%%%%%%%%%%%
 \section{Discussion}
 \label{Disc}
 In this work, we have established the resemblance of codeword organization between the retinal network and the Hopfield network. In
previous studies, the memory retrieval function of Hopfield network
was empirically compared to the behavior of real
networks~\cite{Amit-1985,Amit-1987,Tkacik-2014,Ganmor-2015}. However, no theoretical framework was proposed to
build a solid relationship between these artificial and real biological networks. In fact, they are naturally distinct in terms of detailed
parameters. Surprisingly, we have found that the two networks both similarly organize their codewords. The
clustering of codewords has been identified by the first-order phase transition in the codeword-distance. This
transition is accompanied by hysteresis loops, which becomes
increasingly complex as the network size grows. We have also revealed
that the AS state has a distinct role from other codewords. The number
of codewords surrounding the AS state always grows much more rapidly
as a function of distance compared to that surrounding other codewords. Interestingly,
despite the presence of entropy gaps, most codewords even far away
from the AS state could still have easy access to it because of
their surrounding dense cores of codewords typically extending beyond
the AS state. Thus, the most frequently observed AS state plays a key
role in serving as a hub facilitating neural exploration of the codeword-space.

The only
knowledge a neuronal population can have comes from the population activity of interacting neurons. As shown in our
study, there exists well-designed structure of codewords in the neural state space. The codewords are partitioned into multiple clusters separated by entropy gaps. Moreover, this emergent property remains
even if one-fourth of our data is used to learn the model (see supplementary Fig.~\ref{overf}). Thus the revealed organization structure is most likely an intrinsic property of
the retinal network, and downstream brain areas may benefit from this structure for decoding purpose.

The clustering is functionally advantageous and intimately related to the network function, i.e., pattern completion (error-correction) and 
pattern separation (discrimination ability). 
 Upon repeated presentations of the same visual stimulus, the neural responses show strong trial-to-trial variability~\cite{Tkacik-2014}.
 However, all codewords belonging to the same core perhaps encode the same feature of the semantic information~\cite{Ganmor-2015}.
 This property also allows the neural code to be robust against the ubiquitous noise in nervous systems~\cite{noise-2008}. In an analogous way to error-correcting codes~\cite{Sourlas-1989,Huang-2009pre}, even if the 
 neural codeword is corrupted by a small amount of noise, the dense core structure still allows population coding of stimulus features.
 Therefore, the non-trivial internal structure of the neural codeword-space is useful for the neural population not only to discriminate different
 neural activity pattern, but also to carry out error-correction~\cite{Markus-2008,Fiete-2011}.

 The retina as an early visual system should adapt to the visual
stimulus distribution to efficiently transmit relevant information to
downstream brain areas. The energy landscape shaped by the neural
interactions likely depends on the natural scene statistics. It is
therefore interesting to study their relationship under the current
context. 

The codeword-space structure quantitatively predicts the fluctuation plateau of the simulated 
neural dynamics starting from LEM. Hence, our analytic framework establishes the relationship between the simulated neural dynamics and 
clustering of codewords. In previous studies, the match between spontaneous neural activity and the stimulus-evoked activity increases during development especially
for natural stimuli~\cite{Berkes-2011}, and the spontaneous activity outlines the regime of evoked neural responses~\cite{Luczak-2009}. 
Our analysis might
further reveal how spontaneous neural activity is related to the vocabulary of neural codewords a neural circuit learns to internally represent external worlds.

Overall, our study provides an important step to understand the stationary distribution of neural spiking patterns and its
functional relevance, which also sheds light on future studies of the sensory processing in other brain areas.

%%%%%%%%%%%%%%%%%%%%%%%%%%%%%%%%%%%%%%%%%%%%%%%%%%%%%%%
%\section*{Acknowledgments}

\begin{acknowledgments}
We are grateful to Michael J. Berry for sharing us the retinal data. This work was supported by the program for Brain Mapping by Integrated Neurotechnologies 
for Disease Studies (Brain/MINDS) from Japan Agency for Medical Research and development, AMED.
\end{acknowledgments}

%\section*{Contributions}

%HH and TT conceived and designed the study. HH performed data analysis and modeling. HH and TT wrote the manuscript. All authors discussed the results and implications.

%\section*{Competing financial interests}

%The authors declare no competing financial interests.

%%%%%%%%%%%%%%%%%%%%%%%%%%%%%%%%%%%%%%%%%%%%%%%%%%%%%%%%%%%%%%%
\setcounter{figure}{0}    
\renewcommand{\thefigure}{S\arabic{figure}}
\renewcommand\theequation{S\arabic{equation}}
\setcounter{equation}{0}

%\begin{widetext}
\appendix
%\newpage
%\onecolumngrid
%\newpage
%\appendix
%%%%%%%%%%%%%%%%%%%%%%%%%%%%%%%%%%%%%%%%%%%%%%%%%%%%%%%%%%%%%%%%%%
\section*{Methods}
\label{methods}
\subsection{Simultaneous recordings of neural activity in populations of retinal ganglion cells}
The spiking activity of $160$ retinal ganglion cells was collected from a $450\times450$ $\mu m$ patch of the salamander retina,
when a repeated naturalistic movie was presented. The visual stimulus consists of $297$ repeats of a $19$s long movie clip being
a gray movie of swimming fish and swaying water plants in a tank (data courtesy of Michael J. Berry II, see experimental details in the original paper~\cite{Marre-2012,Tkacik-2014}). The spike train data is binned with the 
bin size $\tau=20ms$ reflecting the temporal correlation time scale, yielding about $280\times 10^3$ binary neural codewords for 
model analysis.

\subsection{Maximum entropy model}
For a neuronal population of size $N$, the neural spike trains of duration $T$ are binned at temporal resolution $\tau$, producing $M=\lceil T/\tau\rceil$ samples of
$N$-dimensional binary neural codewords. We use $\sigma_i=+1$ to indicate spiking activity of neuron $i$, and $\sigma_i=-1$ for silent activity.
The neural responses to repeated stimulus are highly variable (so-called trial-to-trial variability, see Fig.~\ref{EvoD}).
To model the neural codeword statistics, we assign each codeword $\boldsymbol{\sigma}$ a cost function (energy in statistical physics jargon) $E(\boldsymbol{\sigma})$, then 
the probability of observing one codeword $\boldsymbol{\sigma}$ is written as $P(\boldsymbol{\sigma})\propto\exp(-E(\boldsymbol{\sigma}))$, where
\begin{equation}\label{energyIsing}
E(\boldsymbol{\sigma})=-\sum_{i}h_i\sigma_i-\sum_{i<j}J_{ij}\sigma_i\sigma_j.
\end{equation}
The spiking bias $h_i$ and neuronal coupling $J_{ij}$ are constructed from the spike train data such that the spiking rate $m_i$
and the pairwise correlation $C_{ij}$ under the model match those computed from the data. High
energy state $\boldsymbol{\sigma}$ corresponds to low probability of observation. This is a low dimensional representation of the original
high dimensional neural codewords, since we need only $N+N(N-1)/2$ model parameters.

To find the model parameters, we apply the maximum likelihood learning principle corresponding to maximizing
the log-likelihood $P(\boldsymbol{\sigma})$ with respect to the parameters. The learning equation is given by
\begin{subequations}\label{LE}
\begin{align}
h_i^{t+1}&=h_i^{t}+\eta\Biggl(\Bigl<\sigma_i\Bigr>_{{\rm data}}-\Bigl<\sigma_i\Bigr>_{{\rm model}}\Biggr),\\
J_{ij}^{t+1}&=J_{ij}^{t}+\eta\Biggl(\Bigl<\sigma_i\sigma_j\Bigr>_{{\rm data}}-\Bigl<\sigma_i\sigma_j\Bigr>_{{\rm model}}\Biggr),
\end{align}
\end{subequations}
where $t$ and $\eta$ denote the learning step and learning rate, respectively. The maximum likelihood learning shown here has a simple 
interpretation of minimizing the Kullback-Leibler divergence between the empirical probability and the model probability~\cite{CM-12,Huang-2013epjb}.
In the learning equation (Eq.~(\ref{LE})), the data dependent terms can be easily computed from the binned neural data.
However, the model expectation of the spiking rate (magnetization in statistical physics) and pairwise correlation is quite hard to evaluate without
any approximations. Here we propose the mean field method to tackle this difficulty. 
 
The statistical properties of the model (Eq.~(\ref{energyIsing})) can be analyzed by the cavity method in the mean field theory~\cite{cavity-2001}.
The self-consistent equations are written in the form of message passing (detailed derivation is given in Refs~\cite{Huang-2009pre,MM-2009}) as
\begin{subequations}\label{bp}
\begin{align}
m_{i\rightarrow a}&=\tanh\left(h_i+\sum_{b\in\partial i\backslash a}\tanh^{-1}\hat{m}_{b\rightarrow i}\right),\\
\hat{m}_{b\rightarrow i}&=\tanh\Gamma_b\prod_{j\in\partial b\backslash i}m_{j\rightarrow b},
\end{align}
\end{subequations}
where $\partial b\backslash i$ denotes the member of interaction $b$ expect $i$, and $\partial i\backslash a$ denotes the interaction set
$i$ is involved in with $a$ removed. $\Gamma_a\equiv J_{ij}$ and $a\equiv (ij)$. $m_{i\rightarrow a}$ is interpreted as the message passing from the neuron $i$ to the interaction $a$ it
participates in, while $\hat{m}_{b\rightarrow i}$ is interpreted as the message passing from the interaction $b$ to its member $i$. Iteration of 
the message passing equation on the inferred model would converge to a fixed point corresponding to a global (local) minimum of the free
energy (in the cavity method approximation~\cite{MM-2009})
\begin{equation}\label{freeE}
F\equiv-\ln Z=-\sum_{i}\ln Z_i+\sum_{a}(|\partial a|-1)\ln Z_a,
\end{equation}
where $Z$ is the normalization constant (partition function) of the model probability $P(\boldsymbol{\sigma})$. The free energy contribution of one neuron 
$Z_i=\sum_{x=\pm1}\mathcal{H}_i(x)$ where $\mathcal{H}_i(x)\equiv e^{xh_i}\prod_{b\in\partial i}\cosh\Gamma_b(1+x\hat{m}_{b\rightarrow i})$, and the free energy contribution of one interaction $Z_a=\cosh\Gamma_a\left(1+\tanh\Gamma_a\prod_{i\in\partial a}m_{i\rightarrow a}\right)$.
 At the same time, the model spiking rate and multi-neuron
correlation can also be estimated as
\begin{subequations}\label{magcorre}
\begin{align}
m_{i}&=\tanh\left(h_i+\sum_{b\in\partial i}\tanh^{-1}\hat{m}_{b\rightarrow i}\right),\\
C_a&=\frac{\tanh\Gamma_a+\prod_{i\in\partial a}m_{i\rightarrow a}}{1+\tanh\Gamma_a\prod_{i\in\partial a}m_{i\rightarrow a}}.
\end{align}
\end{subequations}
We have defined $m_i=\left<\sigma_i\right>$ and $C_a=\left<\prod_{i\in\partial a}\sigma_i\right>$. Note that the iteration converges in a few steps at each learning stage, 
and estimated magnetizations as well as correlations are used in the gradient ascent learning step. Here the multi-neuron correlation is calculated directly from the cavity method approximation~\cite{Zecchina-2011} and expected to be accurate enough for current 
neural data analysis. Another advantage is the low computational cost. A more accurate expression could be derived from linear response theory~\cite{Huang-2012pre}
with much more expensive computational cost.

Finally, one can also estimate the entropy of the model from the fixed point of the message passing equation. The entropy is defined as
$S=-\sum_{\boldsymbol{\sigma}}P(\boldsymbol{\sigma})\ln P(\boldsymbol{\sigma})$, and it measures the capacity of the neural population for information transmission.
More obvious variability of the neural responses implies larger entropy value. Based on the standard thermodynamic relation, $S=-F+E$, where
$E$ is the energy of the neural population and given by
\begin{subequations}\label{Energ}
\begin{align}
E&=-\sum_{i}\Delta E_i+\sum_{a}(|\partial a|-1)\Delta E_a,\\
\Delta E_i&=\frac{h_i\sum_{x=\pm1}x\mathcal{H}_i(x)+\sum_{x=\pm1}\mathcal{G}_i(x)}{\sum_{x=\pm1}\mathcal{H}_i(x)},\\
\Delta E_a&=\Gamma_a\frac{\tanh\Gamma_a+\prod_{i\in\partial a}m_{i\rightarrow a}}{1+\tanh\Gamma_a\prod_{i\in\partial a}m_{i\rightarrow a}},\\
\begin{split}
\mathcal{G}_i(x)&=\sum_{b\in\partial i}e^{xh_i}\left[\Gamma_b\sinh\Gamma_b(1+x\hat{m}_{b\rightarrow i})+x\Gamma_b\cosh\Gamma_b(1-\tanh^{2}\Gamma_b)\prod_{j\in\partial b\backslash i}m_{j\rightarrow b}
\right]\\
&\times\prod_{a\in\partial i\backslash b}\cosh\Gamma_a(1+x\hat{m}_{a\rightarrow i}).
\end{split}
\end{align}
\end{subequations}

\subsection{Distance-constrained entropy analysis}
To uncover the internal structure of the neural codeword-space, we introduce a modified probability measure~\cite{Huang-JPA2013} 
\begin{equation}\label{energyIsingprob}
P(\boldsymbol{\sigma})=\frac{1}{Z}\exp\left(\sum_{i}\beta h_i\sigma_i+\sum_{i<j}\beta J_{ij}\sigma_i\sigma_j+x\sum_{i}\sigma_i^{*}\sigma_i\right),
\end{equation}
where $\beta$ is the inverse temperature or neural reliability, and the coupling field $x$ is introduced to control the overlap between the neural codeword $\boldsymbol{\sigma}$ and a reference one $\boldsymbol{\sigma}^{*}$.

The partition function $Z$ can be approximated by a saddle point analysis, i.e., $Z\simeq\exp(Ns(d,\epsilon)-\beta N\epsilon+xNq)$, from which the 
free energy per neuron $f$ (density) is given by $-\beta f=s(d,\epsilon)-\beta\epsilon+xq$, where $\epsilon$ is the energy density ($E/N$), $s(d,\epsilon)$ the entropy density ($S/N$) and 
$q$ the typical value of the overlap ($\boldsymbol{\sigma}^{T}\boldsymbol{\sigma}^{*}/N$). Note that the Hamming distance per neuron is related to the overlap by $d=(1-q)/2$. According to the double Legendre transform, the entropy density is calculated via $s(d,\epsilon)=-\beta f+\beta\epsilon-xq$. $e^{Ns(d,\epsilon)}$ counts the number of valid configurations 
around the reference satisfying both the distance constraint ($d$) and the energy density ($\epsilon$). Here $\beta$ controls the energy level and $x$ selects the overlap or Hamming distance. The overlap $q$ is given by $q=\frac{1}{N}\sum_i\sigma_i^{*}m_i$ with $m_i$ being calculated under
the modified probability measure. $(x,\beta)$ obeys the following equations: $\partial s(d,\epsilon)/\partial d=2x$ 
and $\partial s(d,\epsilon)/\partial \epsilon=\beta$. In this setting, the above iteration equations (Eq.~(\ref{bp})) remain unchanged except that the bias is 
changed to $h_i\rightarrow\beta h_i+x\sigma_i^{*}$ and the coupling is rescaled as $J_{ij}\rightarrow\beta J_{ij}$. For the real neuronal network,
the neural reliability $\beta=1$, since the constructed biases and couplings reflect the neural noise observed in the spike train data.
For the Hopfield model, higher $\beta$ implies weaker thermal fluctuation and may correspond to a retrieval phase for pattern completion. 

Note that to compute the entropy curve for metastable or unstable branches of distance-coupling field curve, one has to fix $d$ by searching for
compatible coupling field $x$, e.g., by the secant method~\cite{Numer-2006}.

\subsection{Finding local energy minima from neural activity pattern}
To search for a local energy minimum starting from any given neural activity pattern, we use greedy descent dynamics (GDD) in the 
energy landscape~\cite{Tkacik-2014}. To be more precise, for each neuron, we flip its activity if the flip will decrease the energy. If we could not
decrease the energy by flipping any neuron's activity, then a local energy minimum is identified. Such minima are also called single-flip stable 
attractors, i.e., their energy can not be decreased by flipping any single neuron's activity. We choose randomly a pattern set of size $2000$
from the neural data to ensure that any two patterns are rarely identical. By applying the GDD method, we identify a LEM set whose size is much smaller than that of the pattern set, with
a large portion of patterns evolving to the all-silent state. The number of LEM increases with the network size.
These LEM are then expressed in a low dimensional space (called multidimensional scaling analysis (MDS)~\cite{MDS-2008}).
MDS represents the proximity between LEM in the high dimensional space with some degree of fidelity by the distance
between points in the low dimensional space.

\subsection{Independent maximum entropy model}
In the case of fitting only the first moments (mean spiking activity), the distance entropy can be computed exactly. The 
result is given by 
\begin{equation}\label{indcode}
s(q(x))=\frac{1}{N}\sum_i\ln2\cosh(h_i+x\sigma_i^{*})-\frac{1}{N}\sum_i(h_i+x\sigma_i^{*})\tanh(h_i+x\sigma_i^{*}),
\end{equation}
where $h_i=\frac{1}{2}\ln\frac{1+m_i}{1-m_i}$ and $q(x)=\frac{1}{N}\sum_i\sigma_i^{*}\tanh(h_i+x\sigma_i^{*})$. 

\section*{Supplementary figures}
\begin{figure}
\centering
  (a)        \includegraphics[bb=0 0 364 265,scale=0.6]{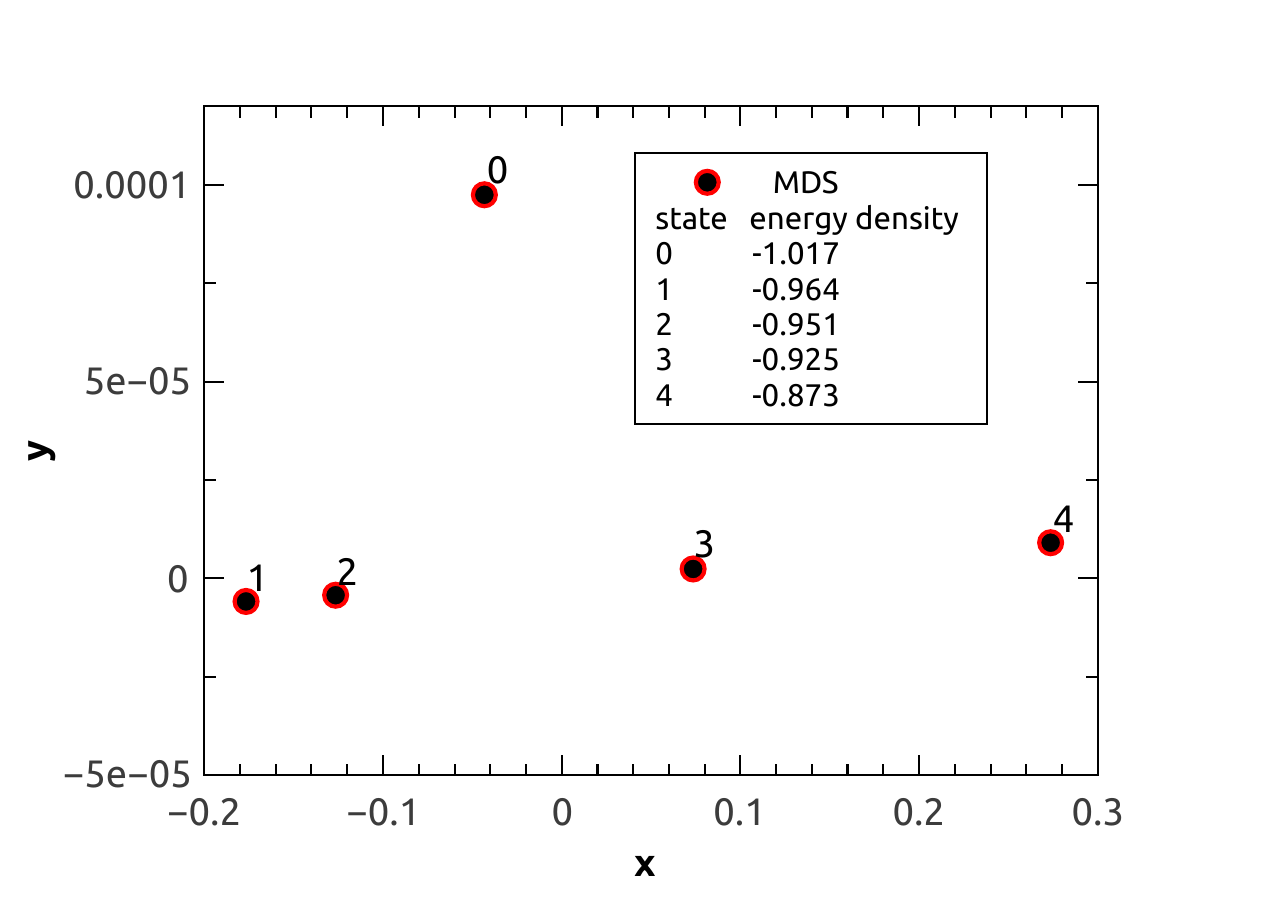}
           \hskip .1cm
 (b)    \includegraphics[bb=0 0 545 380,scale=0.45]{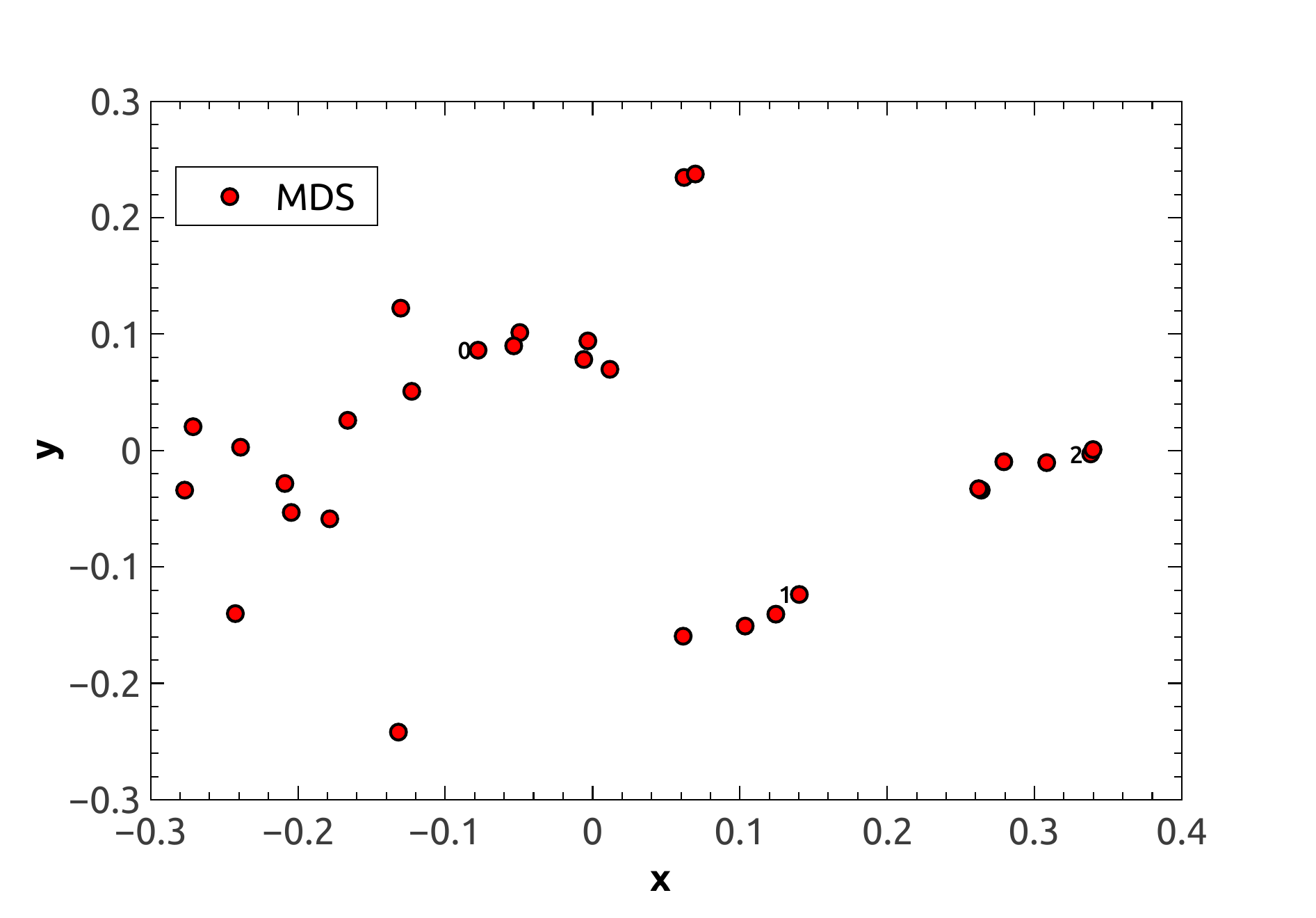}
  \caption{(Color online)
  Low dimensional representation of LEM (identified by GDD) by multidimensional scaling (MDS) analysis ($y,x$ serve as
  coordinates). State $0$ is the AS state. (a) MDS map for a population of $60$ neurons (see Fig.~\ref{retina}).
  (b) MDS map for a population of $100$ neurons (see Fig.~\ref{retinaLN}), in which $29$ LEM are identified by GDD method. It becomes difficult to
  represent faithfully these LEM in a low dimensional space (some information are lost), nevertheless, the map still shows how they
  are distributed.
     }\label{mdsLN}
 \end{figure}

\begin{figure}[h!]
(a)    \includegraphics[bb=0 0 351 265,scale=0.65]{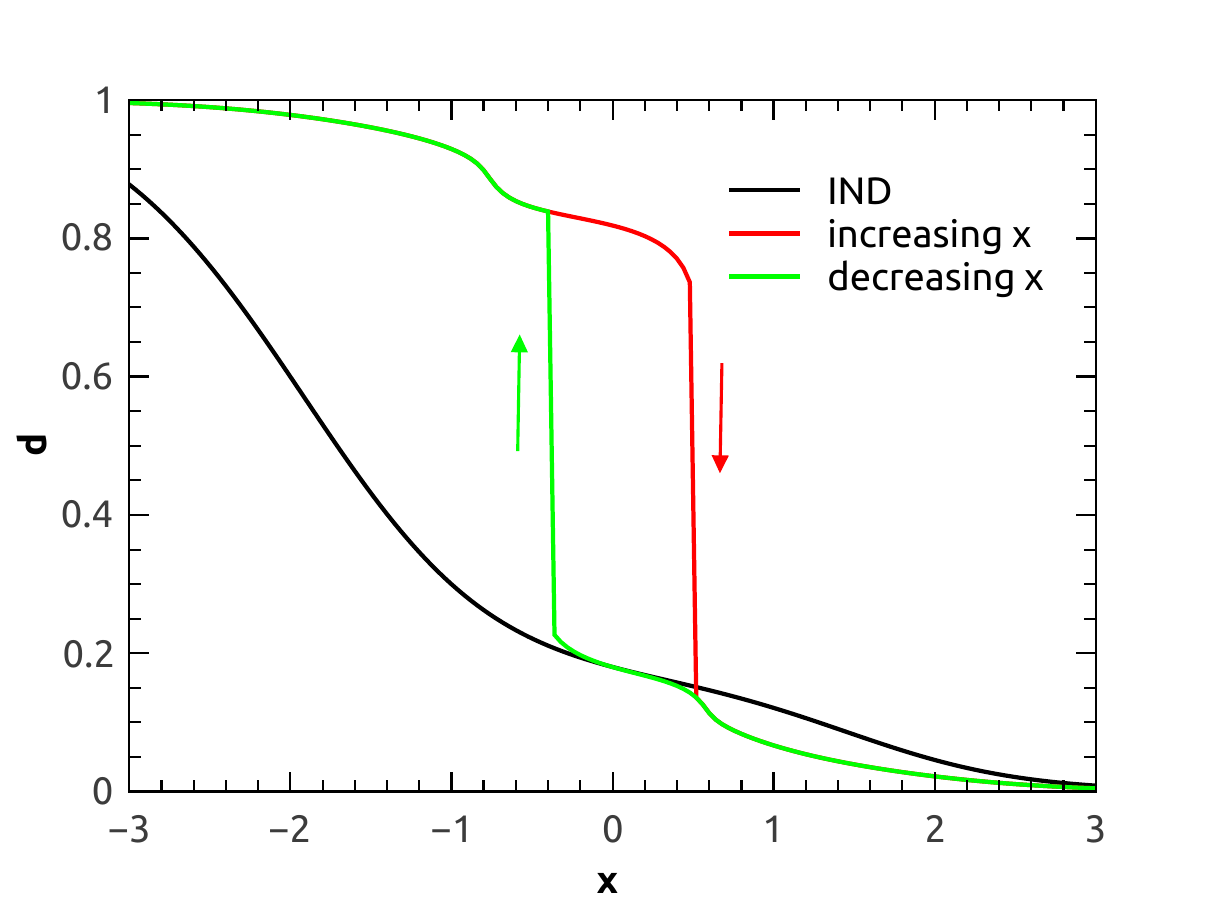}
     \hskip .1cm
 (b)    \includegraphics[bb=0 0 352 265,scale=0.65]{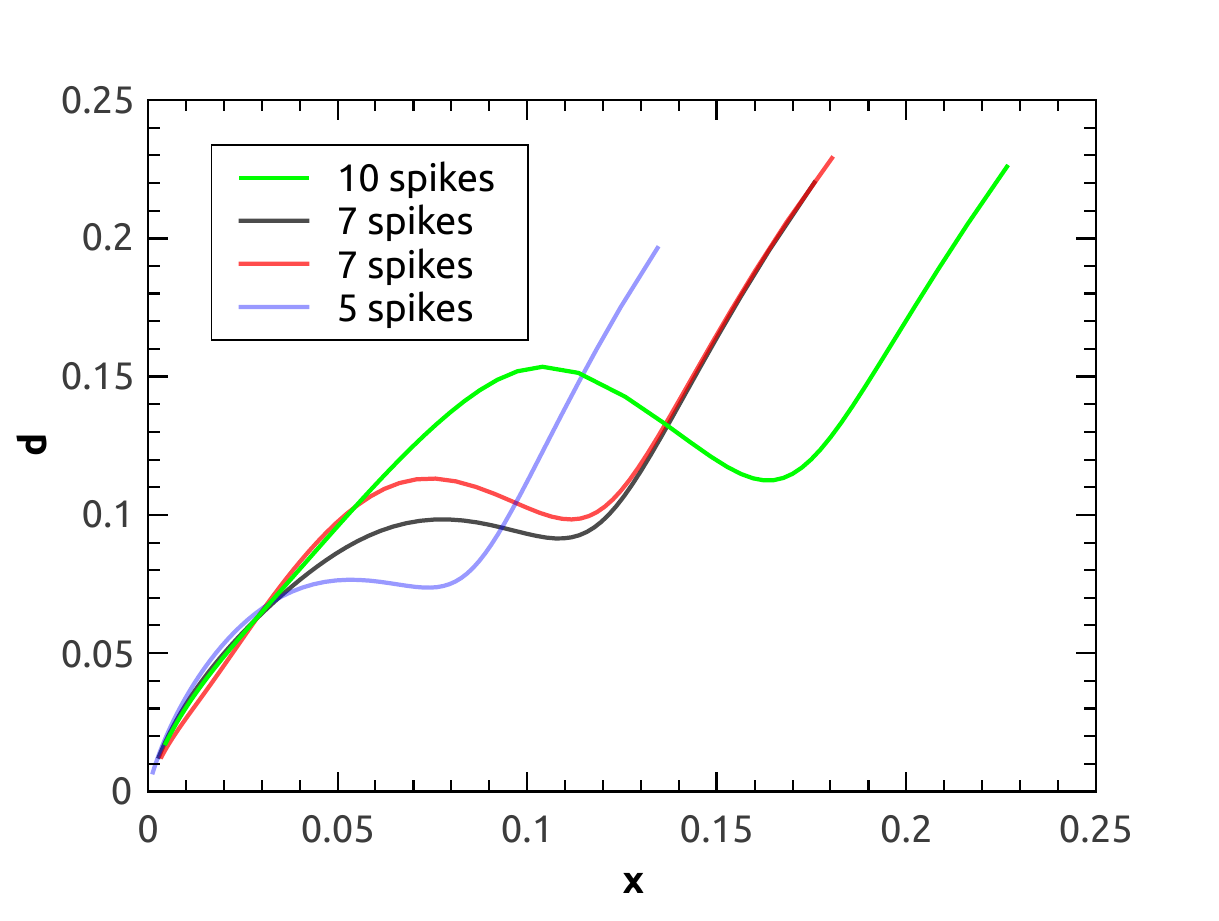}
     \vskip .1cm
     (c)    \includegraphics[bb=0 0 352 265,scale=0.62]{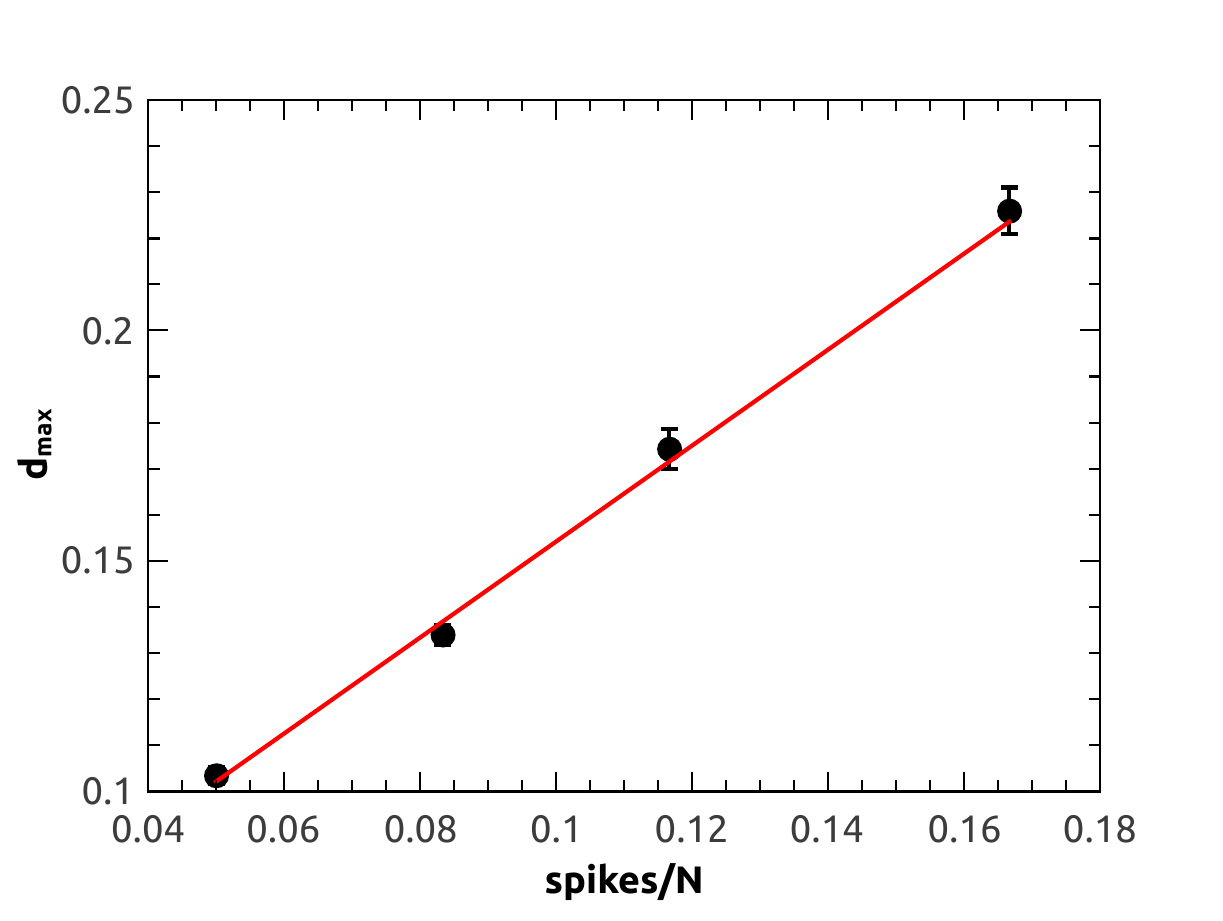}
     \hskip .1cm
 (d)    \includegraphics[bb=0 0 473 375,scale=0.45]{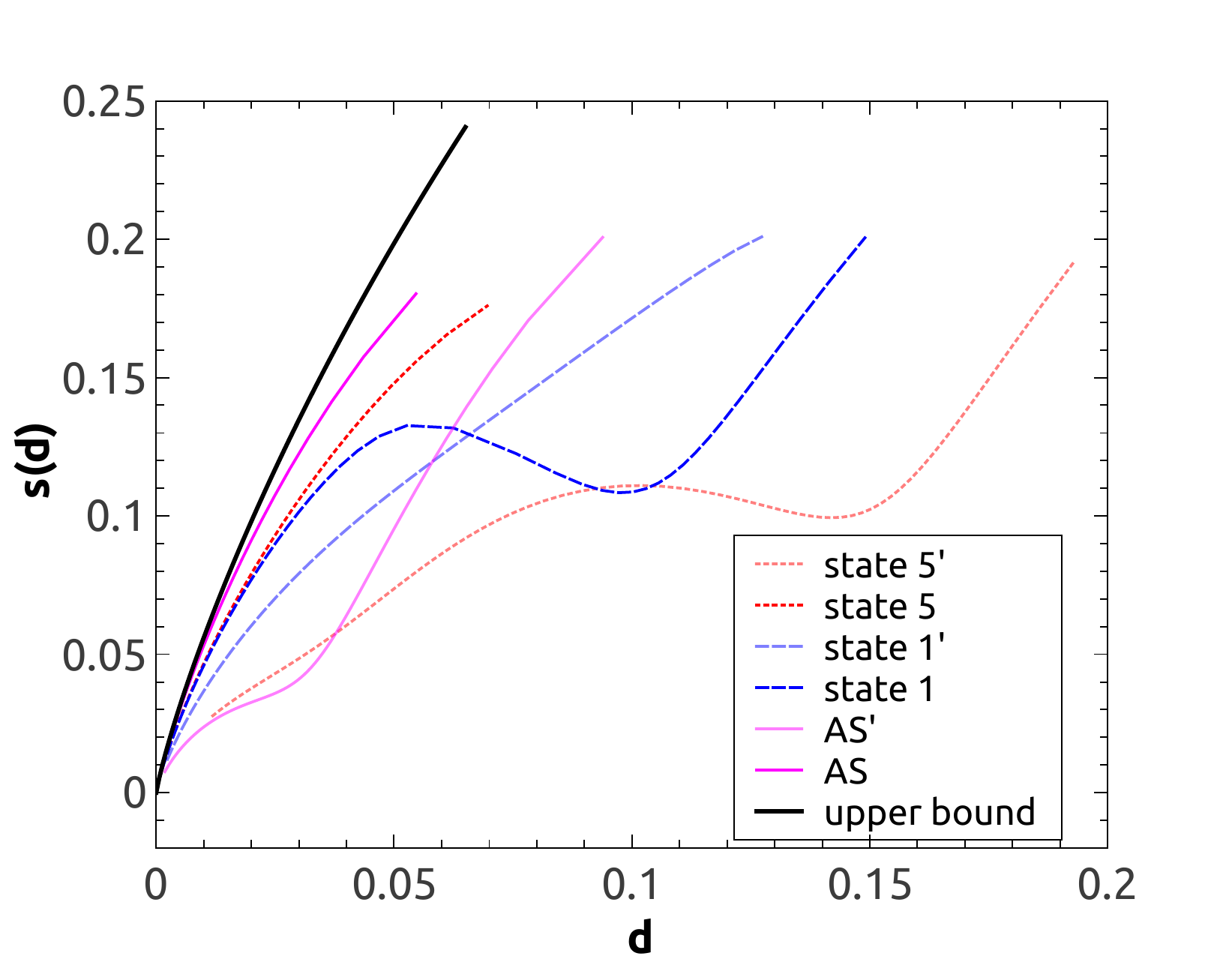}
   \vskip .1cm
 (e)    \includegraphics[bb=0 0 365 265,scale=0.6]{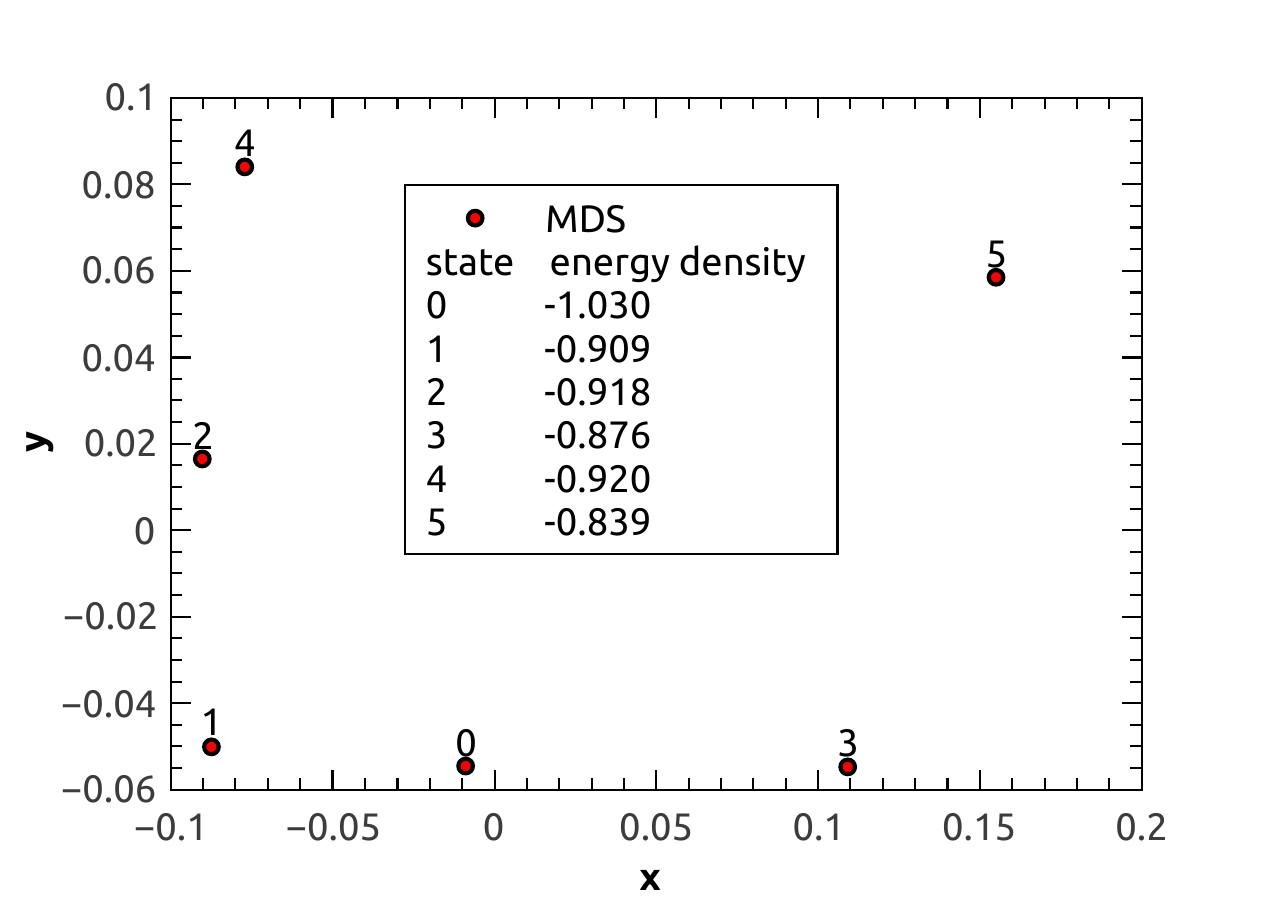}
     \vskip .1cm
  \caption{(Color online) Entropy landscape of the neural data ($N=60$, another typical example).
  (a) a first order phase transition in Hamming distance when the coupling field is tuned. The transition disappears for an independent model (IND). (b) distance
  entropy from reference neural codewords of different spike-counts. (c) maximum distance $d_{{\rm max}}$ at which the low-$d$ branch in the hysteresis loop terminates
  versus spike-counts of the reference (distance from AS state). Five references for each spike-count are considered. The line is a linear fit (slope=$1.041\pm0.040$). (d) distance entropy from neural codewords and their corresponding 
  LEM. (e) low dimensional representation of LEM corresponding to (d) by multidimensional scaling (MDS) analysis ($y,x$ serve as
  coordinates). State $0$ is the AS state. 
     }\label{retina02}
 \end{figure}

\begin{figure}
  (a) \includegraphics[bb=0 0 364 265,scale=0.65]{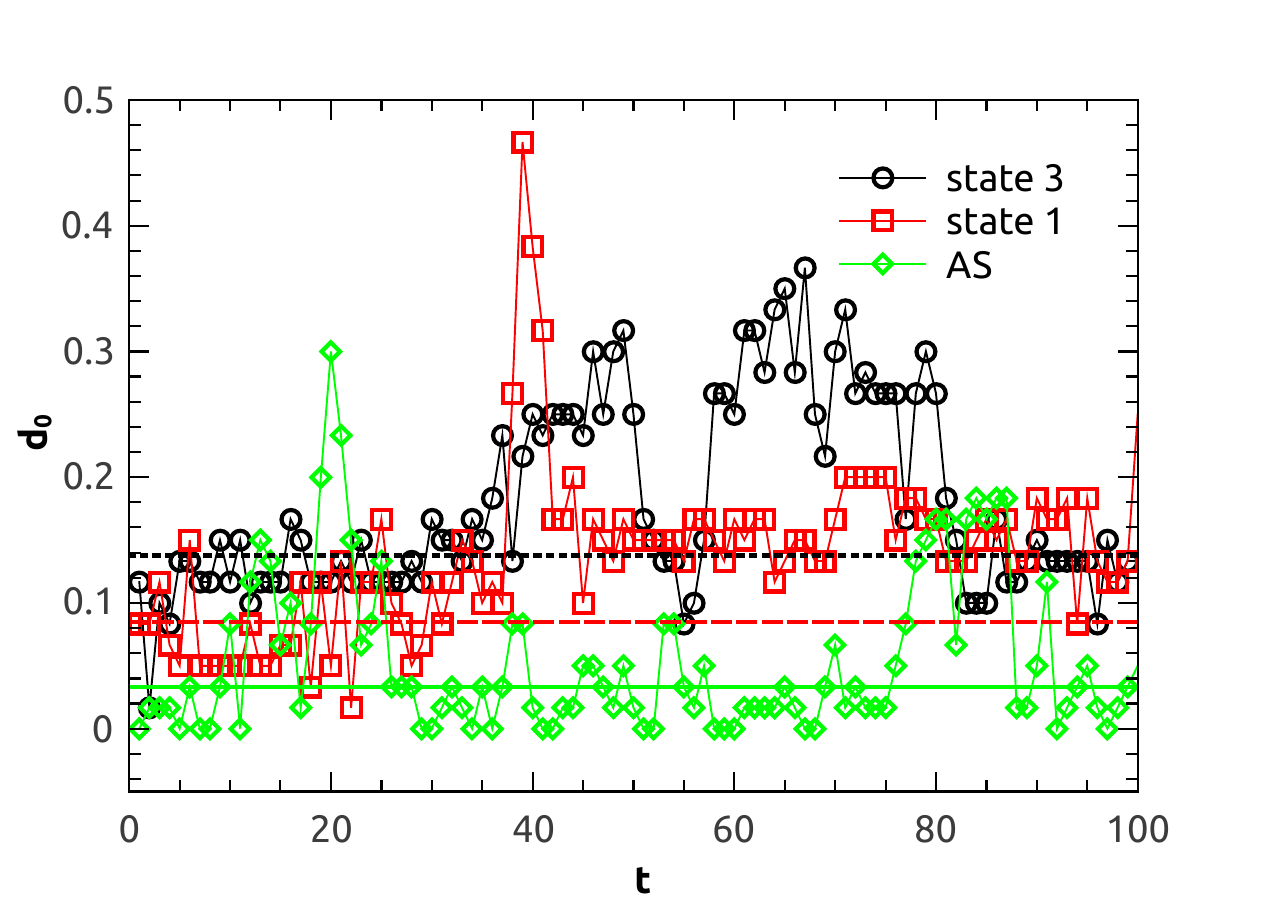}
   \hskip .1cm
    (b) \includegraphics[bb=0 0 351 265,scale=0.65]{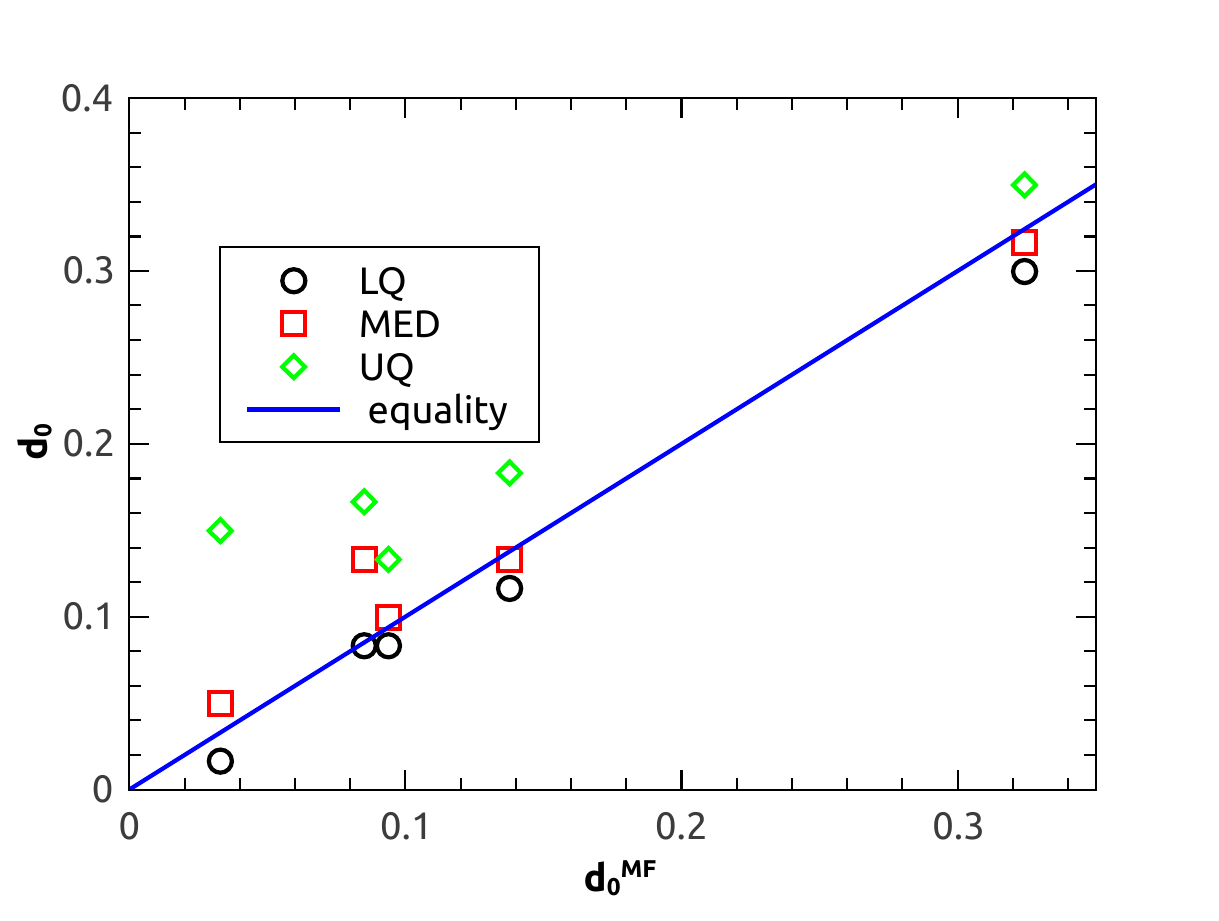}
 \vskip .1cm
 (c) \includegraphics[bb=0 0 488 355,scale=0.5]{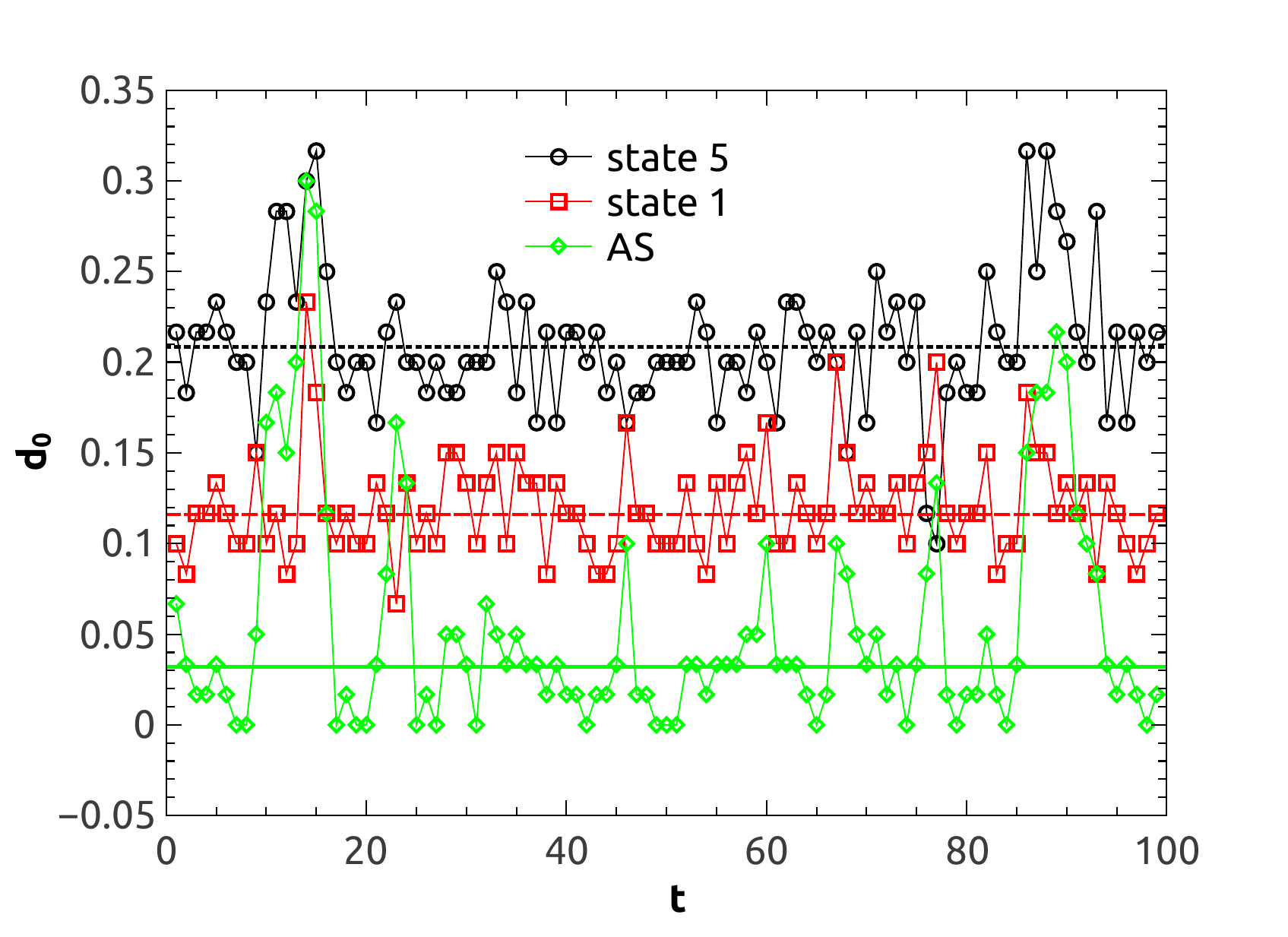}
   \hskip .1cm
    (d) \includegraphics[bb=0 0 351 265,scale=0.65]{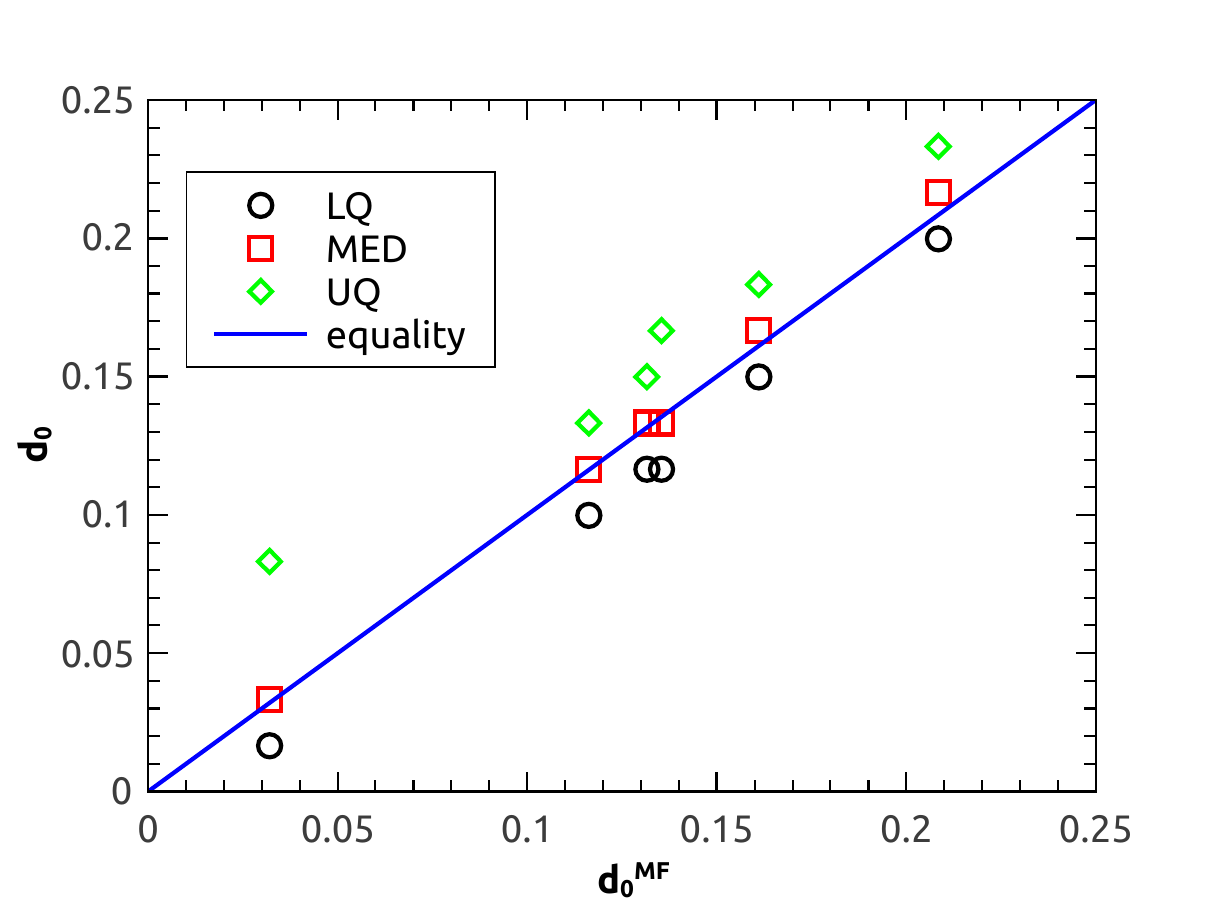}
 \vskip .1cm
  \caption{(Color online)
  Neural dynamics starting from LEM ($d_{0}(t)$). (a,b) the network with $60$ neurons in the main text. (c,d) the network with $60$ neurons corresponding to 
  Fig.~\ref{retina02}. (a,c) typical trajectory observed in simulations. The (solid, dashed, dotted) line is
  the theoretical prediction computed at $x=0$ for each reference. (b,d) the fluctuation plateau of $d_{0}(t)$ is predicted by the mean field theory ($d_{0}^{{\rm MF}}$).
  Five trials from the 
  same reference are considered for each data point. Each trial lasts for $100$ steps. LQ: lower quartile; MED: median; UQ: upper
  quartile. 
     }\label{dynn60}
 \end{figure}

\begin{figure}
          \includegraphics[bb=0 0 352 265,scale=0.65]{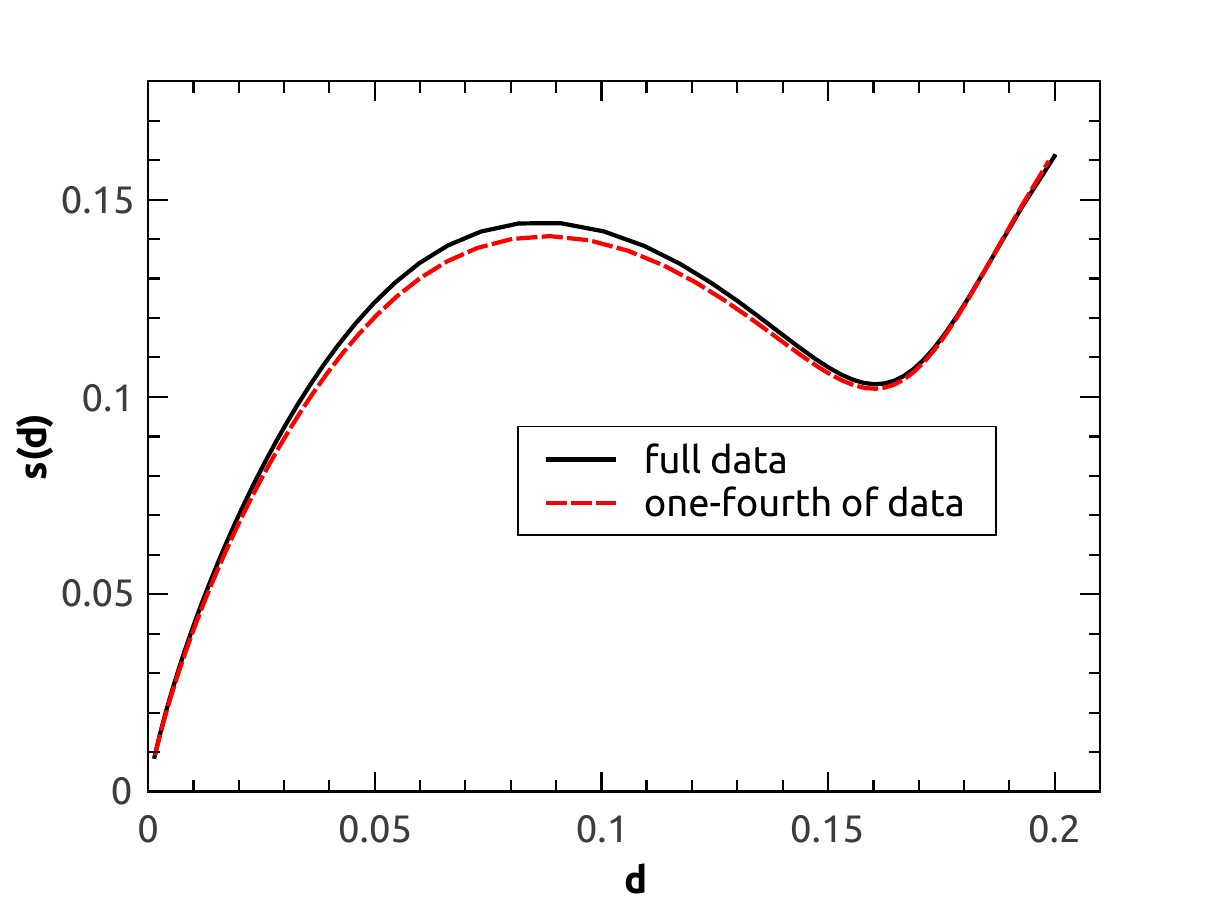}
  \caption{(Color online)
  Entropy landscape for $N=60$. The same reference codeword has $10$ spikes but two data of different lengths are learned.
     }\label{overf}
 \end{figure}

\begin{figure}[h!]
(a)    \includegraphics[bb=0 0 726 401,scale=0.55]{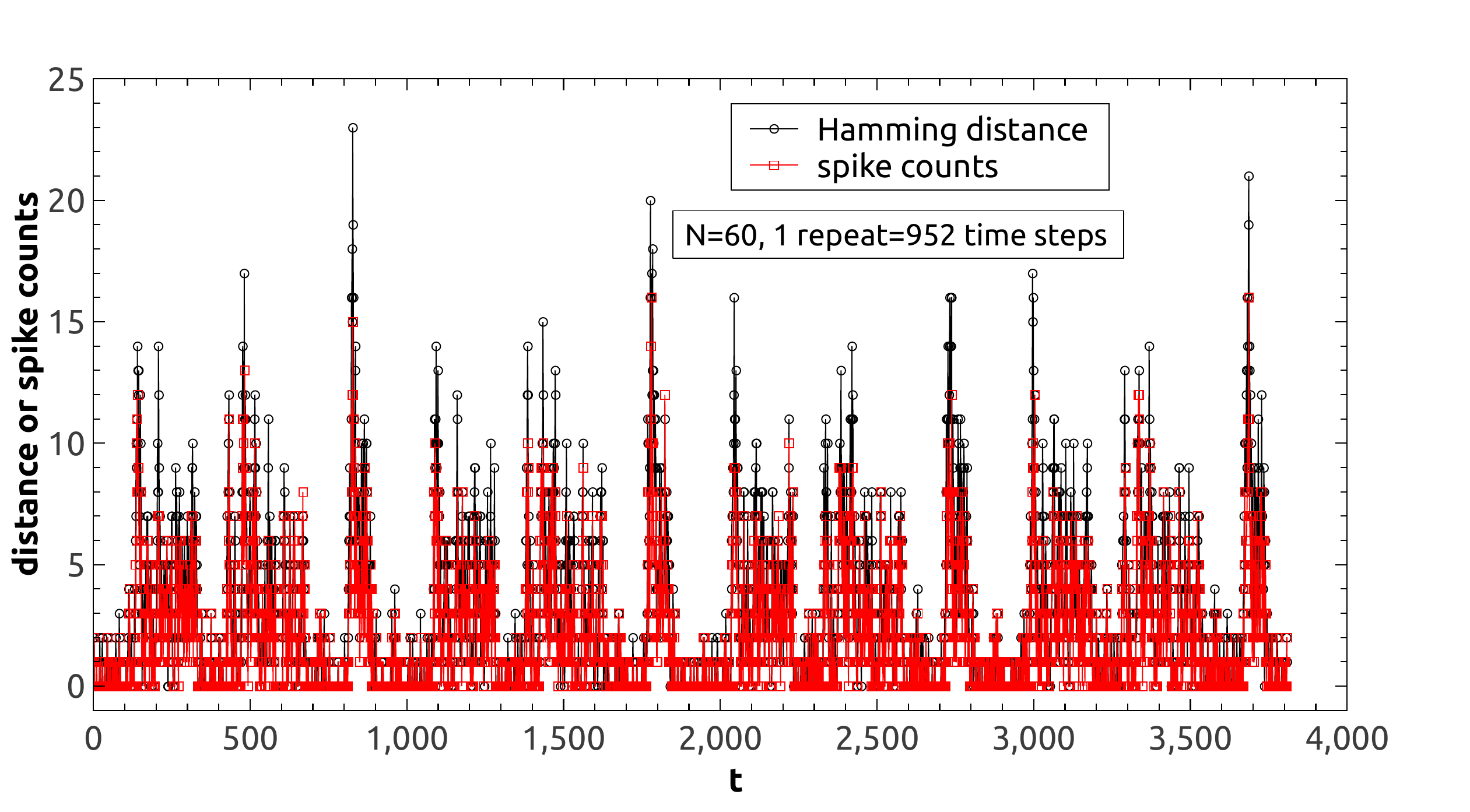}
     \vskip .1cm
 (b)    \includegraphics[bb=0 0 726 400,scale=0.55]{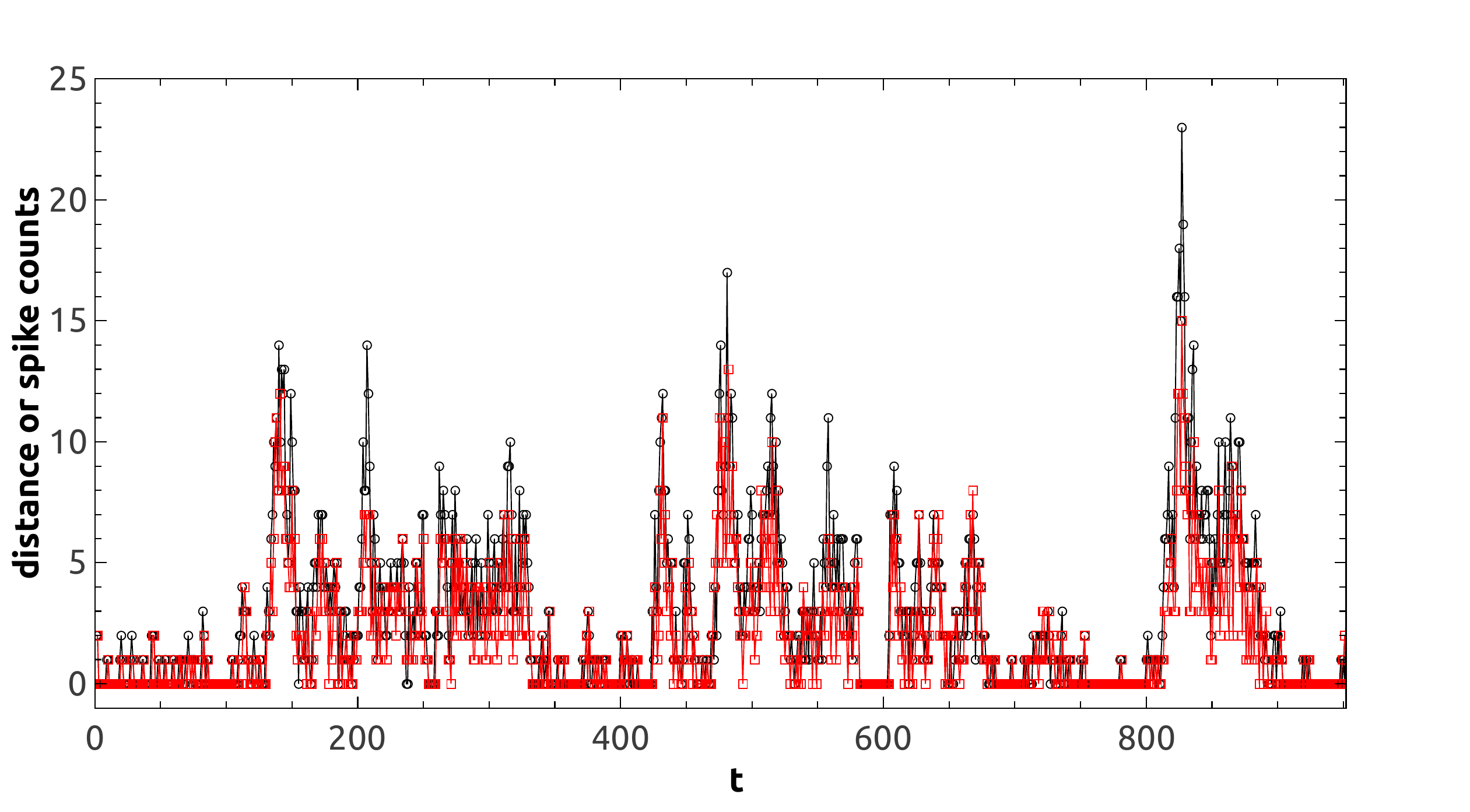}
     \vskip .1cm
  \caption{(Color online) Distance or spike-counts evolution of the neural data ($N=60$, typical example shown in the main text).
  (a) the profile for four repeats. Trial-to-trial variability is observed. (b) the profile for only one repeat. The AS state is frequently visited, 
  and the neural network seems to explore the state space by local moves.
     }\label{EvoD}
\end{figure}

Fig.~\ref{mdsLN} corresponds to Fig.~\ref{retina} and Fig.~\ref{retinaLN} in the main text. Fig.~\ref{retina02} shows another typical example of 
a network of $60$ neurons. The qualitative properties do not change. Fig.~\ref{dynn60} shows the neural dynamics result for smaller networks ($N=60$). 
Fig.~\ref{overf} shows that the problem structure is 
not affected by the finite sampling of the data. Fig.~\ref{EvoD} shows the role of the AS state with temporal information included.
The time-dependent Hamming distance is defined as $d(t)\equiv\frac{N-\sum_i\sigma_i(t)\sigma_i(t+1)}{2}$, and the time-dependent spike-counts $r(t)\equiv\sum_i\delta(\sigma_i(t)-1)$.
 
%%%%%%%%%%%%%%%%%%%%%%%%%%%%%%%%%%%%%%%%%%%%%%%%%%%%%%%%%%%%%%%%%%%%%
%\bibliography{ref}

\end{document}